\documentclass[12pt]{elsarticle}
\usepackage{etex}

\usepackage{xcolor}
\usepackage{relsize}

\usepackage{anyfontsize}
\usepackage{todonotes}

\usepackage{amssymb}
\usepackage{amsmath}
\usepackage{subfigure}
\usepackage{amsmath}
\usepackage{oz}
\usepackage{refcal}
\usepackage{url}
\usepackage[strings]{underscore}

\usepackage{rcp}

\usepackage{time}
\usepackage{rcoms}
\usepackage{BT}
\usepackage{counters}
\usepackage{opsem}
\usepackage{multicol}

\usepackage{wmm}
\definecolor{dkgreen}{rgb}{0,0.6,0}
\definecolor{gray}{rgb}{0.5,0.5,0.5}
\definecolor{mauve}{rgb}{0.58,0,0.82}

\newcommand{\LOCALSWord}{\mathop{\mathbf{lcl}}}
\newcommand{\IFWord}{\mathop{\mathbf{if}}}
\newcommand{\THENWord}{\mathop{\mathbf{then}}}
\newcommand{\ELSEWord}{\mathop{\mathbf{else}}}
\newcommand{\WHILEWord}{\mathop{\mathbf{while}}}
\newcommand{\DOWord}{\mathop{\mathbf{do}}}

\newcommand{\wro}{\mathrel{\overset{\textsc{w}}{\Leftarrow}}}
\newcommand{\canReorder}[2]{#1 \wro #2}
\newcommand{\flush}{flush}

\def\HC{\cite{HerdingCats}{} }

\renewcommand{\refrule}[1]{Rule~(\ref{rule:#1})}
\renewcommand{\reflaw}[1]{Law~(\ref{law:#1})}
\newcommand{\reflaws}[2]{Laws~(\ref{law:#1}) and~(\ref{law:#2})}
\newcommand{\refrulea}[1]{Rule~(\ref{rule:#1}a)}
\newcommand{\refruleb}[1]{Rule~(\ref{rule:#1}b)}

\newcommand{\smallasgn}{\asgnsmall}

\renewcommand{\reg}[2]{\mathtt{REG~#1, #2}}

\renewcommand{\refeqn}[1]{(\ref{eq:#1})}
\renewcommand{\refeqns}[2]{(\ref{eq:#1},\ref{eq:#2})}
\renewcommand{\WHbc}{\While b~\Do~c}
\renewcommand{\reffig}[1]{Fig.~\ref{fig:#1}}

\newcommand{\eieio}{\keywordfont{eieio}}

\newcommand{\scat}{\mathbin{{}^\mathsmaller{\mathsmaller{\frown}}}}

\renewcommand{\store}[2]{#1 \asgn #2}
\renewcommand{\storex}[1]{\store{x}{#1}}

\newcommand{\varDistinct}[2]{#1 \overset{\mathsf{v}}{\nsim} #2}
\newcommand{\vnfi}[2]{\varDistinct{#1}{#2}}
\newcommand{\vnfix}[1]{\vnfi{x}{#1}}
\newcommand{\vnfiy}[1]{\vnfi{y}{#1}}
\newcommand{\vnfir}[1]{\vnfi{r}{#1}}
\newcommand{\vnfirb}{\vnfir{b}}

\newcommand{\vnfixf}{\vnfix{f}}
\newcommand{\vnfiye}{\vnfiy{e}}

\newcommand{\loadDistinct}[2]{#1 \overset{\mathsf{sv}}{\nsim} #2}
\newcommand{\loadDistinctef}{\loadDistinct{e}{f}}

\def\atomicsep{~,~}
\newcommand{\PIDSet}{\textsc{PID}}

\newcommand{\addrShift}[2]{#2_{\mathsf{\mathsmaller{\&\!\mbox{\tiny +}}}#1}}

\newcommand{\localsnp}[2]{\LOCALSWord~#1 @ #2} %np = no parens
\renewcommand{\locals}[2]{(\localsnp{#1}{#2})}
\newcommand{\globals}[2]{(\keywordfont{store}~#1 @ #2)}

\renewcommand{\prefix}[2]{#1 \cbef #2}
\renewcommand{\prefixa}[1]{\prefix{\aca}{#1}}
\renewcommand{\prefixac}{(\prefixa{\cmdc})}
\renewcommand{\prefixacp}{(\prefixa{\cmdc'})}

\renewcommand{\eval}[2]{{#2}_{#1}}
\newcommand{\evalse}{\evals{e}}
\newcommand{\Read}[2]{\guard{#1 = #2}}

\newcommand{\Readxv}{\Read{x}{v}}

\newtheorem{theorem}[equation]{Rule}
\newcommand{\ruledefNamed}[4]{
	\begin{minipage}[b]{#1}
	\begin{theorem}[#2]
		\label{rule:#3}
	~
	\vskip 0.5mm
	$\quad
			#4
	$
	\\
	$~ $
	\\
	$~ $
	\end{theorem}
	\end{minipage}
}

\newcommand{\code}[1]{$#1$}

\newcommand{\seqT}[1]{\langle #1 \rangle}

\newcommand{\sees}[3]{\mathsf{see}_{#1}^{#2}(#3)}

\newcommand{\lwfenced}[1]{\mathsf{lwf}(#1)}
\newcommand{\lwfencedn}{\lwfenced{\pidn}}

\newcommand{\sys}[1]{\defaultFont{#1}}
\newcommand{\syss}{\sys{s}}

\newcommand{\pid}[1]{\textsc{#1}}
\newcommand{\pidn}{\pid{n}}
\newcommand{\pidm}{\pid{m}}
\newcommand{\pidns}{{\cal S}}

\newcommand{\srq}[4]{(#2 \mapsto #3)^{#1}_{#4}}
\newcommand{\srqnxv}[1]{\srq{\pidn}{x}{v}{#1}}
\newcommand{\srqnxvns}{\srqnxv{\pidns}}
\newcommand{\srqnxvn}{\srqnxv{\{\pidn\}}}
\newcommand{\srqmxv}[1]{\srq{\pidm}{x}{v}{#1}}
\newcommand{\srqmxvns}{\srqmxv{\pidns}}

\newcommand{\wseq}{\omega}

\renewcommand{\Update}[3]{{#1}_{[#2 \smallasgn #3]}}
\newcommand{\Repl}[3]{{#1}_{[#2\backslash#3]}}

\renewcommand{\ttdef}{\mathbin{:\!:\!=}}

\renewcommand{\cbar}{\mathbin{~|~}}

\newcommand{\mathttbf}[1]{\mbox{\code{#1}}}

\renewcommand{\If}{\mathrel{\keywordfont{if}}}
\renewcommand{\Then}{\mathrel{\keywordfont{then}}}
\renewcommand{\Else}{\mathrel{\keywordfont{else}}}
\renewcommand{\While}{\mathrel{\keywordfont{while}}}
\renewcommand{\Do}{{\mathrel{\keywordfont{do}}}}

\newcommand{\wpre}[2]{wp(#1, #2)}

\newcommand{\CASWord}{CAS}
\newcommand{\CAS}[3]{CAS(#1, #2, #3)}
\newcommand{\CASxre}{\CAS{x}{r}{e}}

\newcommand{\MT}[1]{$\mathtt{#1}$}

\newcommand{\reorderable}{\mathrel{\overset{\textsc{r}}{\Leftarrow}}}

\newcommand{\reordOf}[2]{#1 ~\overset{\textsc{r}}{\refsto}~ #2}

\newcommand{\reordOfccp}{\reordOf{\cmdc}{\cmdc'}}

\newcommand{\nreorderable}{\not\!\reorderable}
\newcommand{\ro}{\reorderable}
\newcommand{\nro}{\nreorderable}

\newcommand{\loadgate}{\keywordfont{loadgate}}
\newcommand{\storegate}{\keywordfont{storegate}}

\newcommand{\loadgatelbl}{\keywordfont{loadgate}}
\newcommand{\loadlbl}[2]{#1 \asgnsmall #2}

\renewcommand\notin{\not\,\!\in}

\newcommand{\ltif}[1]{\quad \mbox{if #1}}

\newcommand{\fwd}[2]{{#2}_{\langle #1 \rangle}}
\newcommand{\fwdab}{\fwd{\acb}{\aca}}

\newcommand{\storage}[2]{(\keywordfont{storage}~ #1 @ #2)}
\newcommand{\thread}[2]{(\keywordfont{tid}_{#1} ~ #2)}

\newcommand{\befSkip}{}
\newcommand{\cbefSkip}{}

\newcommand{\defaultFont}[1]{#1}

\newcommand{\cmd}[1]{\defaultFont{#1}}
\newcommand{\cmdc}{\cmd{c}}
\newcommand{\cmdd}{\cmd{d}}

\newcommand{\ac}[1]{\defaultFont{#1}}
\newcommand{\aca}{\ac{\alpha}}
\newcommand{\acb}{\ac{\beta}}

\newcommand{\pr}[1]{\defaultFont{#1}}
\newcommand{\prp}{\pr{p}}
\newcommand{\cbef}{\mathbin{\mathtt{;}}}
\newcommand{\bef}{\,\centerdot\,}

\renewcommand{\asgn}{\mathbin{\mathtt{:\!=}}}

\renewcommand{\RegStoreWord}{{\mathttbf{MOV}}}
\renewcommand{\LoadWord}{{\mathttbf{LDR}}}
\renewcommand{\StoreWord}{{\mathttbf{STR}}}
\renewcommand{\RMWWord}{{\mathttbf{rmw}}}

\renewcommand{\RegStoreWord}{{\mathttbf{reg}}}
\renewcommand{\LoadWord}{{\mathttbf{load}}}
\renewcommand{\StoreWord}{{\mathttbf{store}}}
\renewcommand{\RMWWord}{{\mathttbf{rmw}}}

\renewcommand{\regstore}[2]{\mathtt{\RegStoreWord~#1 , #2}}
\renewcommand{\load}[2]{\mathtt{\LoadWord~#1 , #2}}
\renewcommand{\store}[2]{\mathtt{\StoreWord~#1 , #2}}
\renewcommand{\rmw}[3]{\mathtt{\RMWWord~#1,#2,#3}}
\newcommand{\guard}[1]{\GuardWord~#1}
\newcommand{\guarde}{\guard{e}}
\newcommand{\guardb}{\guard{b}}

\def\keywordfont{\mathbf}

\renewcommand{\guard}[1]{[#1]}
\renewcommand{\fence}{\keywordfont{fence}}
\renewcommand{\cfence}{\keywordfont{cfence}}
\newcommand{\lfence}{\keywordfont{lwfence}}
\newcommand{\wwfence}{\keywordfont{fence.st}}

\newcommand{\guardlbl}[1]{[#1]}
\newcommand{\guardlble}{\guardlbl{e}}

\newcommand{\fencelbl}{\keywordfont{fence}}

\newcommand{\asgnlbl}{\mathrel{\,\mathtt{:=}\,}}

\renewcommand{\prefix}[2]{#1 \cbef #2}
\renewcommand{\prefixa}[1]{\prefix{\aca}{#1}}
\renewcommand{\prefixac}{(\prefixa{\cmdc})}
\renewcommand{\prefixacp}{(\prefixa{\cmdc'})}

\journal{Theoretical Computer Science}

\begin{document}

%\lstset{language=IMP}

\begin{frontmatter}

%\title{Implications of weak memory models on imperative programming languages}
%\title{Higher-level programming in the presence of relaxed memory models}
%\title{Foundations for developing hardware weak-memory model code from specifications}
\title{A high-level operational semantics for hardware weak memory models}

\author{Robert J. Colvin}
\ead{r.colvin@uq.edu.au}

\address{
	%\institution{
	School of Electrical Engineering and Information Technology\\
	%} \department{
	The University of Queensland
	%}
}

\author{Graeme Smith}
\ead{smith@itee.uq.edu.au}

\address{
	%\institution{
	School of Electrical Engineering and Information Technology\\
	%}
	%\department{
	The University of Queensland
	%}
}

\begin{abstract}
Modern processors deploy a variety of weak memory models, which for efficiency reasons may execute instructions in an order different to
that specified by the program text.  The consequences of instruction reordering can be complex and subtle, and can impact on ensuring
correctness.
In this paper we build on extensive work elucidating the semantics of assembler-level languages on hardware architectures with weak memory models (specifically TSO,
ARM and POWER) and lift the principles to a straightforward operational semantics which allows reasoning at a higher level of abstraction.  To this end we
introduce a wide-spectrum language that encompasses operations on abstract data types as well as low-level assembler code, define its operational semantics using a
novel approach to allowing reordering of instructions, and derive some refinement laws that can be used to explain behaviours of real
processors.  In this framework memory models are mostly distinguished via a pair-wise static ordering on instruction types that determines when later
instructions may be reordered before earlier instructions. In addition, memory models may use different types of storage systems. For instance,
non-multicopy atomic systems allow sibling processes to see updates to different variables in different orders.
%Our goal is
%to support reasoning about implementations of data structures for modern processors with respect to an abstract specification.  

We encode the semantics in the rewriting engine Maude as a model-checking tool, and develop confidence in our framework by validating our semantics against
existing sets of \textit{litmus tests} -- small assembler programs -- comparing our results with those observed on hardware and in existing semantics.  We also use the
tool as a prototype to model check implementations of data structures from the literature against their abstract specifications.

\OMIT{
Modern processors deploy a variety of weak memory models, which for efficiency reasons may execute instructions in an order different to that specified by the
program text.  The consequences of instruction reordering can be complex and subtle, and can impact on developing correct software.  In this paper we develop a
general operational semantics for high-level programs running on weak memory models as a basis for developing verification tools and techniques.  Building on
earlier work, the semantics succinctly captures how instructions in a programming language can be pair-wise reordered.
%especially with respect to updates to the global state.

The semantics is encoded in a tool, which we use to validate instantiations of the semantics for the ARM and POWER processors against an
established set of assembly-level litmus tests for these processors. We also apply the tool as a prototype model checker to establish
that a simple lock algorithm provides mutual exclusion, and that implementations of a stack and double-ended queue satisfy their abstract specifications. Finally,
we derive some properties from the operational semantics that support algebraic and refinement-based reasoning as a basis for future theorem proving support.
}

\end{abstract}

%\keywords{Weak memory models, semantics, verification}

\begin{keyword}
weak memory models \sep operational semantics \sep verification
\end{keyword}

\end{frontmatter}

\section{Introduction}
Modern processor architectures provide a challenge for developing efficient and correct software.  Performance can be improved by parallelising
computation and utilising multiple cores, but communication between threads is notoriously error prone.  Weak memory models go further and improve
overall system efficiency through sophisticated techniques for batching read and writes to the same variables and to and from the same
processors.  However, code that is run on such memory models is not guaranteed to execute in the order specified in the program text, creating
unexpected behaviours for those who are not forewarned \citep{AdveBoehm2010}.  To aid the programmer, architectures typically provide memory barrier/fence instructions which can enforce
thread-local order corresponding to the program text, but if overused fences can eliminate 
performance gains.

Previous work on formalising hardware weak memory models has resulted in abstract formalisations which were developed incrementally through
communication with processor vendors and rigorous testing on real machines \citep{UnderstandingPOWER,HerdingCats,ModellingARMv8}.  A large
collection of ``litmus tests'' \citep{LitmusTests,Mador-Haim2010} demonstrate the sometimes confusing behaviour of hardware.  We build on this
existing work and provide a programming language and operational semantics that runs on the same relaxed principles that apply to the assembler
instructions.  The semantics is validated against litmus tests from the literature, and then applied to model check some realistic concurrent data structures. 

%Our theory is encoded in Maude \citep{Maude}, which provides an efficient rewriting engine.  
%The operational semantics generates a trace consisting of a sequence of actions which may be loads, stores, fences, etc.
%The behaviour of an architecture is specified as the relationship
%between individual types of actions, in addition to the behaviour of the storage system.

We begin in \refsect{overview} with the basis of a straightforward operational semantics that allows reordering of instructions according to pair-wise
relationships between instructions, and an overview of the results of the paper.  
In \refsect{syntax} we introduce our wide-spectrum language and an informal description of the instructions.
In \refsect{semantics} we
describe the semantics in more detail, and derive some properties that support algebraic and refinement-based reasoning as a basis for theorem proving.
Later
we show the instantiations of the thread-local definitions to three well-known
weak memory models, TSO \citep{x86-TSO} in \refsect{tso}, ARM \citep{ModellingARMv8} in \refsects{armv8}{arm} and POWER \citep{UnderstandingPOWER} in \refsect{power}.  We then consider the implications of
weak memory models on more complex algorithms in \refsect{higher-level-code}: we verify a simple lock
\citep[Sect. 7.3]{HerlihyShavit2011}, the Treiber lock-free stack \citep{Treiber86} running on ARM and POWER, and find (and fix) a bug in an implementation of the
Chase-Lev work-stealing deque (double-ended queue) \citep{ChaseLev05} developed specifically for ARM \citep{LeWorkStealingPPoPP13}.  
We discuss related work in \refsect{related-work}.

\paragraph{Contributions}
This paper extends our earlier work \cite{FM18} in the following ways:
\begin{itemize}
\item
We address TSO.
\item
We take into account a more recent version of ARM.
\item
We compare our semantics to a larger set of litmus test results (over 18,000 in this paper vs. approx 1,000 in \cite{FM18}), and as a result handle more
constructs (e.g., POWER's lightweight fences and $\eieio$ fences), and other types of constraints (e.g., address shifting).
\item
We apply the semantics to more case studies.
\end{itemize}

\section{Instruction reordering in weak memory models}
\label{overview}

\subsection{Thread-local reorderings}

%Imperative languages explicitly build an order into the program text -- the ``program order'' -- via sequential composition.  
It is typically 
assumed processes are executed in a fixed sequential order (as given by sequential composition -- the ``program order'').  
However program order may be inefficient, 
e.g., when retrieving the value of a variable from main memory after setting its value, as in \MT{x \asgn 1 \cbef r \asgn x}, and hence
weak memory models sometimes allow execution to appear out of program order to improve overall system efficiency.  Specifically, in the above case, the
value 1 may be used for $r$ in later calculations, possibly including writing to some other shared variable, without waiting for the update to $x$ to
propagate to all other threads in the system.
While many reorderings can seem surprising, there are basic principles at play which limit the number of possible permutations, the key
being that the new ordering of instructions preserves the original sequential intention.

A classic example of weak memory models producing unexpected behaviour is the ``store buffer'' pattern below \cite{LitmusTests}.
Assume that all variables are initially 0, that $r_1$ and $r_2$ are thread-local variables, and that $x$ and $y$ are shared variables.
\begin{equation}
	(x \asgn 1 \cbef r_1 \asgn y)
	\pl 
	(y \asgn 1 \cbef r_2 \asgn x)
\end{equation}
It is possible to reach a final state in which $r_1 = r_2 = 0$ in several weak memory models:  the two assignments in each
process are independent (they reference different variables), and hence can be reordered.  From a sequential semantics perspective,
reordering the assignments in process 1, for example, preserves the final values
for $r_1$ and $x$.
%We analyse this store buffer behaviour in a formal context in \refsect{MP-refinement}.

Assume that $c$ and $c'$ are programs represented as sequences of atomic actions
$\aca \cbef \acb \cbef \ldots$, as in a sequence of instructions of a thread or more abstractly a semantic trace.  
Program $c$ may be reordered to $c'$, written $\reordOfccp$, if the following holds:
\begin{enumerate}
\item
$\cmdc'$ is a permutation of the actions of $\cmdc$, possibly with some modifications due to \emph{forwarding} (see below).

\item
$\cmdc'$ preserves the sequential semantics of $\cmdc$.  For example, in a weakest preconditions semantics \cite{GuardedCommands},
for all predicates $P$,
$\wpre{\cmdc}{P} \imp \wpre{\cmdc'}{P}$.
%$(\all S @ \wpre{\cmdc}{S} \imp \wpre{\cmdc'}{S})$.

\item
$\cmdc'$ preserves \emph{coherence-per-location} with respect to $\cmdc$
(cf. \T{po-loc} in \cite{HerdingCats}).
This means that
the order of updates and accesses of each shared variable, considered individually, is maintained.
%Below we preserve coherence-per-location by considering pairs of actions.
\end{enumerate}
%Condition 1 emerges straightforwardly from processor documentation, where new instructions are not inserted at run time (though some may effectively be eliminated by
%forwarding).  Condition 2 is reverse engineered from the reordering constraints on different processor types: we have not seen this explicitly stated anywhere, and is not
%immediately obvious in the interactions of low-level instruction types, but it appears to be sensible and justifies why some reorderings are allowed and not others.
%Condition 3 is described in the literature (e.g., \cite{HerdingCats}) and is similar to Condition 2, but on a global level.
We formalise these constraints in the context of pair-wise reordering of instructions below.
The key challenge for reasoning about programs executed on a weak memory model is that
the behaviour of $\cmdc \pl \cmdd$ is in general quite different to the behaviour of $\cmdc' \pl \cmdd$, even if $\reordOfccp$.
%We focus in this paper on the principles for ARM and POWER processors; for space reasons we do not address TSO \cite{x86-TSO}, which has fewer relevant instruction types
%(e.g., only one type of fence) and stricter conditions on reordering.

\subsection{Reordering and forwarding instructions}
\label{overview-reordering}

We write $\alpha \ro \beta$ if instruction $\beta$ may be reordered before instruction $\alpha$.
For TSO the well-known weakening of instruction order is that loads can appear before stores (to different variables).
If we let $x$ and $y$ be shared variables, $r$ be a local variable, and $v$ some value, we can represent this as
\[
	x \asgn v \ro r \asgn y
\]
where $x$ and $y$ are distinct.  This rule applies to a specific case of assignment statements that correspond to assembler-level stores and loads; the
relation is generalised to all assignments for TSO in \refsect{tso}. %\reffig{tso-ro}.

We give the more complex rule for reordering of updates in the ARM and POWER memory models below.
We use the notation $\varDistinct{e}{f}$ to mean that expressions $e$ and $f$ do not reference any variables in common; 
hence $\vnfixf$ can be read as ``$x$ is not free in $f$''.
The related notation $\loadDistinct{e}{f}$ is weaker, requiring only that the \emph{shared} variables of $e$ and $f$ are distinct.
%Let $\vnfixf$ mean that $x$ does not appear free in the expression $f$,
%and say expressions $e$ and $f$ are \emph{load-distinct} if they do not reference any common shared variables.
\begin{eqnarray}
	%\vnfixf &iff& \mbox{variable $x$ is not free in expression $e$}
	\vnfi{e}{f} &iff& \mbox{the free variables of $e$ and $f$ are distinct}
	\\
	\loadDistinctef &iff&
		\mbox{the \emph{shared} variables of $e$ and $f$ are distinct}
	\\
	x \asgn e \ro y \asgn f
	%\quad
	&iff&
	%\quad
	%\parbox{0.65\textwidth}{
		%\emph{if:} 
		%(i) $x$, $y$ are distinct; 
	\mbox{%
		(i) $\varDistinct{x}{y}$
		~(ii) $\vnfixf$
		%\\
		~(iii) $\vnfiye$
		%and 
		~(iv) $\loadDistinctef$ 
	}
\label{eq:reordering-principle}
\end{eqnarray}
Note that $\ro$ as defined above is symmetric, however when calculated after the effect of forwarding is applied (as described below) there are instructions 
that may be reordered in one direction but not the other.  The relation is neither reflexive nor transitive.

Provisos (i), (ii) and (iii) ensure executing the two assignments in either order results in the same final values for $x$ and $y$, and proviso (iv)
maintains order on accesses of the shared state.
If two updates do not refer to any common variables they may be reordered.  

Proviso (i) eliminates reorderings such as
$
	\reordOf{
		(x \asgn 1 \cbef x \asgn 2)
	}{
		(x \asgn 2 \cbef x \asgn 1)
	}
$ 
which would violate the sequential semantics (the final value of $x$).
Proviso (ii) eliminates reorderings such as
$
	\reordOf{
		(x \asgn 1 \cbef r \asgn x)
	}{
		(r \asgn x \cbef x \asgn 1)
	}
$
which again would violate the sequential semantics (the final value of $r$).
Proviso (iii) eliminates reorderings such as
$
	\reordOf{
		(r \asgn y \cbef y \asgn 1)
	}{
		(y \asgn 1 \cbef r \asgn y)
	}
$
which again would violate the sequential semantics (the final value of $r$).
Proviso (iv), requiring the update expressions' shared variables are distinct, preserves coherence-per-location,
eliminating reorderings such as
$
	\reordOf{
		(r_1 \asgn x \cbef r_2 \asgn x)
	}{
		(r_2 \asgn x \cbef r_1 \asgn x)
	}
$, where $r_2$ may receive an earlier value of $x$ than $r_1$ in an environment which modifies $x$.

\renewcommand{\LoadWord}{{\tt LDR}}
\renewcommand{\StoreWord}{{\tt {STR}}}
The instructions used in the above examples, where each instruction references at most one global variable and uses simple integer values, correspond to the basic 
load and store instruction types of ARM and POWER processors.  We may instantiate \refeqn{reordering-principle} to such instructions, giving reordering rules such as the
following, which states that a store may be reordered before a load if they are to different locations
($r_1 \asgn y \ro x \asgn r_2$).
We use ARM syntax to emphasise the application to a real
architecture.
\begin{equation}
	\load{r_1}{y} \ro \storex{r_2}
\end{equation}

In practice, proviso (ii) may be circumvented by \emph{forwarding}.%
\footnote{We adopt the term ``forwarding" from ARM and POWER \cite{HerdingCats}. The equivalent effect is sometimes referred to as \emph{bypassing} on TSO \cite{x86-TSO}.}  
This refers to taking into account the effect of the update moved earlier on the expression of the other update.
We write $\fwd{\alpha}{\beta}$ to represent the effect of forwarding the (assignment) instruction $\alpha$ to the instruction $\beta$.
For assignments we define
\begin{equation}
\label{eq:assignment-forwarding}
	\fwd{x \asgnsmall e}{(y \asgn f)}
	~~=~~
	y \asgn (\Repl{f}{x}{e})
	\quad
	if
	\quad
	\mbox{$e$ does not refer to global variables}
\end{equation}
where
the term $\Repl{f}{x}{e}$ stands for the syntactic replacement in expression $f$ of references to $x$ with $e$.
The proviso of \refeqn{assignment-forwarding} prevents additional loads of globals being introduced by forwarding.

\OMIT{
The final constraint eliminates reorderings such as
$
	\reordOf{
		(r_1 \asgn x \cbef r_2 \asgn r_1)
	}{
		(r_2 \asgn x \cbef r_1 \asgn x)
	}
$.
Note that the assignment to $r_2$ has been affected by forwarding.  Here a second load of $x$ has been introduced, possibly giving
different final values for $r_1$ and $r_2$.  If instead $x$ was a local register, there would be no harm in the reordering (and forwarding)
since the environment may not modify the register.
The proviso does allow reorderings with forwarding such as
$
	\reordOf{(r \asgn 1 \cbef x \asgn r)}{(x \asgn 1 \cbef r \asgn 1)}
$.
}

We specify the reordering and forwarding relationships with other instructions such as branches and fences in
the sections on specific architectures.

\subsection{General operational rules for reordering}
\label{overview-rule}

The key operational principle allowing reordering is given by the following transition rules
for a program $(\aca \cbef \cmdc)$, i.e., a program with initial instruction $\aca$.
\begin{equation}
\label{rule:reorder-rule}
	\prefixac \tra{\aca} \cmdc
	~~~~~~~(a)
	\qquad
	\qquad
	\Rule{
		\cmdc \tra{\acb} \cmdc'
		\quad
		\aca \ro \fwd{\aca}{\acb}
	}{
		\prefixac \ttra{\fwd{\aca}{\acb}} \prefixacp
	}
	~~~~~~~(b)
	%%%\Rule{
		%%%\fwd{\aca}{\cmdc} \tra{\acb} \cmdc'
		%%%\fwd{\aca}{\cmdc} \tra{\acb} \cmdc'
		%%%\quad
		%%%\aca \ro \acb
	%%%}{
		%%%\prefixac \ttra{\acb} \prefixacp
	%%%}
	%%\qquad
	%%\left(
	%%\Rule{
		%%\cmdc \tra{\acb} \cmdc'
		%%\quad
		%%\aca \ro \acb
	%%}{
		%%\prefixac \ttra{\fwd{\aca}{\acb}} \prefixacp
	%%}
	%%\right)
\end{equation}
\refrulea{reorder-rule}
is the straightforward promotion of the first instruction into a step in a trace, similar to the basic prefixing rules of CCS
\cite{CCS} and CSP \cite{CSP}.
\refruleb{reorder-rule},
however, states that, unique to weak memory models, an instruction of $\cmdc$, say $\acb$, can happen before $\aca$, provided that
$\fwd{\aca}{\acb}$ can be reordered before $\aca$
according to the rules of the architecture.
Note that we forward the effect of $\aca$ to $\acb$ before deciding if the reordering is possible.

Applying \refruleb{reorder-rule} then \refrulea{reorder-rule} gives the following reordered behaviour of two assignments.
\begin{equation}
	(r \asgn 1 \cbef x \asgn r \cbef \Skip)
	\ttra{x \asgnsmall 1}
	(r \asgn 1 \cbef \Skip)
	\ttra{r \asgnsmall 1}
	\Skip
\end{equation}
We use the command $\Skip$ to denote termination.
The first transition above is possible because we calculate the effect of $r \asgn 1$ on the update of $x$ before executing that update, i.e.,
$\fwd{r \asgnsmall 1}{x \asgn r} = x \asgn 1$.
	
The definition of instruction reordering, $\aca \ro \acb$
is architecture-specific
(instruction forwarding, $\fwd{\aca}{\acb}$, is constant for the architectures we consider).%
\footnote{
Typically this is the only definition required to specify an architecture's instruction ordering,
but some behaviours may require specialised operational rules, e.g., see \refsect{elimEarlierWrites}.
In addition, different architectures may have different
storage subsystems, however, and these need to be separately defined (see \refsect{storage-subsystem-semantics}).
}

\OMIT{
The instantiations for {\em sequentially consistent\/} processors (i.e., those which do not have a weak memory model) are trivial: $\aca \nro
\acb$ for all $\aca, \acb$, and there is no forwarding.  Since reordering is not possible \refruleb{reorder-rule} never
applies and hence the standard prefixing semantics is maintained.
TSO is relatively straightforward: loads may be reordered before stores (provided they reference different shared variables).
In our framework there is no need to explicitly model local buffers, as the forwarding (bypassing) mechanism ensures that only the most recently
stored value for a global $x$ is used locally (or $x$'s value is retrieved from the storage system).
In this paper we focus on the more complex ARM and POWER memory models. These memory models are very similar, the notable difference being the inclusion of the \emph{lightweight fence}
instruction in POWER. Due to space limitations, we omit lightweight fences in this paper but see the appendix of
\cite{FM18arXiv} for a full definition.
%that has been validated against litmus tests.
}

\subsection{Reasoning about reorderings}

The operational rules allow a standard trace model of correctness to be adopted, i.e., we say program $\cmdc$ 
\emph{refines to} program $\cmdd$, written $\cmdc \refsto \cmdd$, iff every trace of $\cmdd$ is a trace of $\cmdc$.  
Let the program $\aca \bef \cmdc$ have the standard semantics of prefixing, that is, the action $\aca$ always occurs before any action in
$\cmdc$ (\refrulea{reorder-rule}).  Then we can derive the following laws that show the interplay of reordering
and true prefixing.
\begin{eqnarray}
	\aca \cbef \cmdc
	&\refsto&
	\aca \bef \cmdc
	%\aca \cbef \acb \cbef \cmdc
	%&\refsto&
	%\aca \bef \acb \cbef \cmdc
\label{law:keep-order}
	\\
	\aca \cbef (\acb \bef \cmdc)
	&\refsto&
	\fwd{\aca}{\acb} \bef (\aca \cbef \cmdc)
	\qquad
	\mbox{if $\aca \ro \fwd{\aca}{\acb}$}
\label{law:swap-order}
\end{eqnarray}
Note that in \reflaw{swap-order} $\aca$ may be further reordered with instructions in $\cmdc$. Let $\cmdc \pl \cmdd$ denote program $\cmdc$ running concurrently with program $\cmdd$. 
A trace-based interleaving semantics allows us to derive the following laws straightforwardly.
\begin{eqnarray}
	(\aca \bef \cmdc) \pl \cmdd
	&\refsto&
	\aca \bef (\cmdc \pl \cmdd)
	\label{law:fix-interleaving}
	\\
	\cmdc \refsto \cmdc' \land \cmdd \refsto \cmdd'
	&\imp&
	\cmdc \pl \cmdd \refsto \cmdc' \pl \cmdd'
\label{law:mono-pl}
\end{eqnarray}
\reflaw{fix-interleaving} is a typical interleaving law, 
and \reflaw{mono-pl} states that refining either program, or both programs, results in a refinement of their composition. We leave the use of \reflaw{mono-pl}, and properties
such as commutativity, implicit in our
derivations in this paper.

We may use these laws to show how the
``surprise'' behaviour of the store buffer pattern above arises.%
\footnote{
To focus on instruction reorderings we leave local variable declarations and process ids implicit,
and assume a multi-copy atomic storage system (see \refsect{storage-subsystem-semantics}).
}
In derivations such as the following, to save space, we abbreviate a thread $\prefixa{\Skip}$ or $\aca \bef \Skip$ to $\aca$, that is, we omit the trailing $\Skip$.
\begin{derivation}
	\step{
		(x \asgn 1 \cbef r_1 \asgn y \cbefSkip)
		\pl 
		(y \asgn 1 \cbef r_2 \asgn x \cbefSkip)
	}
	\trans{\refsto}{
		From \reflaw{swap-order} (twice), since $x \asgn 1 \ro r_1 \asgn y$ from (\ref{eq:reordering-principle}).
	}
	\step{
		(r_1 \asgn y \bef x \asgn 1 \befSkip)
		\pl 
		(r_2 \asgn x \bef y \asgn 1 \befSkip)
	}
	\trans{\refsto}{
		\reflaw{fix-interleaving} (four times).
	}
	\step{
		r_1 \asgn y \bef r_2 \asgn x \bef x \asgn 1 \bef y \asgn 1 \befSkip
	}
\end{derivation}
If initially $x = y = 0$, a standard sequential semantics shows that $r_1 = r_2 = 0$ is a possible final state in this behaviour.

\section{Wide-spectrum language}
\label{syntax}

In this section we give an overview of the syntax for %and justification of 
our wide-spectrum language. Its elements are actions (instructions) $\alpha$, commands (programs) $c$, processes (local state and a command) $p$, and the top level system $s$,
encompassing a shared state and all processes.
Below $x$ is a variable (shared or local) and $e$ an expression.
\begin{equation}
\label{eq:def-syntax}
\begin{aligned}
	\aca &\ttdef 
		x \asgn e
		\cbar
		\guarde
		\cbar
		\fence
		%\cbar
		%\lfence
		%\cbar
		%\cfence
		\cbar
		\aca^*
	\\
	\cmdc &\ttdef
		\Skip 
		\cbar
		\aca \cbef \cmdc
		\cbar
		\cmdc_1 \choice \cmdc_2
		\cbar
		\WHbc
	\\
	\prp &\ttdef
		(\localsnp{\sigma}{\cmdc})
		\cbar
		\thread{\pidn}{\prp}
		\cbar
		\prp_1 \pl \prp_2
	\\
	\syss &\ttdef
		\globals{\sigma}{\prp}
		%\cbar
		%\storage{\wseq}{\prp}
\end{aligned}
\end{equation}

The basic actions of a weak memory model are
an update $x \asgn e$, a guard $\guarde$, a (full) fence,
%a control fence (see \refsect{armv8})
or a finite sequence of actions, $\aca^*$, executed atomically.  Throughout the paper we denote an empty sequence by
$\eseq$, and construct a non-empty sequence as $\seqT{\alpha_1 \atomicsep \alpha_2 \ldots}$.
ARM and POWER introduce other instruction types, especially different types of fences, which we discuss in the relevant sections.

A command may be the empty command $\Skip$, which is already terminated, a command prefixed by some action $\aca$, a choice between
two commands, or an iteration (for brevity we consider only one type of iteration, the while loop).
\OMIT{
Sequential composition of commands, as opposed to action prefixing, can be defined by induction.
\begin{eqnarray*}
	\Skip \scomp \cmdc = \cmdc
	\qquad
	\qquad
	&&
	\qquad
	\qquad
	(\aca \cbef \cmdc_1) \scomp \cmdc_2 = \aca \cbef (\cmdc_1 \scomp \cmdc_2)
	\\
	(\cmdc_1 \choice \cmdc_2) \scomp \cmdc_3 &=& (\cmdc_1 \scomp \cmdc_3) \choice (\cmdc_2 \scomp \cmdc_3)
\end{eqnarray*}
}

A well-formed process is structured as a process id $\pidn \in \PIDSet$ encompassing a (possibly empty) local state $\sigma$ and command $\cmdc$, i.e., a term
$\thread{\pidn}{\localsnp{\sigma}{\cmdc}}$.
We assume that all local variables referenced in $\cmdc$ are contained in the domain of $\sigma$.

A system is structured as the parallel composition of processes within the global storage system. 
The typical structure is that of
a global
state, $\sigma$, that maps all global variables to their values, which models the storage systems of TSO, the most recent version of ARM \cite{ARMv8.4}, and abstract specifications.
\begin{equation}
\begin{aligned}
\label{eq:system-structure}
	\globals{\sigma}{
		\thread{1}{\localsnp{\sigma_1}{\cmdc_1}} 
		\pl
		\thread{2}{\localsnp{\sigma_2}{\cmdc_2}}
		\pl
		\ldots
	}
	%% \\
	%% \storage{\wseq}{
		%% \thread{1}{\localsnp{\sigma_1}{\cmdc_1}} 
		%% \pl
		%% \thread{2}{\localsnp{\sigma_2}{\cmdc_2}}
		%% \pl
		%% \ldots
	%% }
\end{aligned}
\end{equation}
Older versions of ARM and POWER have a more complex storage system, though the structure of the overall system remains the
same, as discussed in \refsect{storage-subsystem}.
%The storage $\wseq$ injects more nondeterminism into the system than the typical global state approach.

\subsection{Abbreviations}
\label{syntax-abbreviations}

Conditionals are modelled using guards and choice (where false branches are never executed).
\begin{equation}
\label{eq:defn-if}
	\IFbc \sdef (\guard{b} \cbef \cmdc_1) \choice (\guard{\neg b} \cbef \cmdc_2)
\end{equation}
By allowing instructions in $\cmdc_1$ or $\cmdc_2$ to be reordered before the guards one can model \emph{speculative execution}, i.e., early execution of instructions which occur after a branch point \cite{PrimerMemoryConsistency11}: see
\refsect{speculative-execution}.

Although the basic thread language is very simple (reflecting a sequence of instruction on a processor, or a trace in a denotational semantics model) 
we may construct more familiar imperative programming
constructs in the usual way.
Sequential composition of commands, as opposed to action prefixing, can be defined by induction.
\begin{eqnarray}
	\Skip \scomp \cmdc &=& \cmdc
	\\
	(\aca \cbef \cmdc_1) \scomp \cmdc_2 &=& \aca \cbef (\cmdc_1 \scomp \cmdc_2)
	\\
	(\cmdc_1 \choice \cmdc_2) \scomp \cmdc_3 &=& (\cmdc_1 \scomp \cmdc_3) \choice (\cmdc_2 \scomp \cmdc_3)
\end{eqnarray}
Loops are modelled using unfolding, as in \refrule{loop-unfold-rule} below.

Read-modify-write primitives that allow atomic access of more than one variable 
can be modelled as an atomic sequence of steps. For instance,
consider a fenced compare-and-swap (\code{CAS}) instruction, 
where $\CASxre$ updates shared variable $x$ to the value of expression $e$ if $x = r$, and otherwise does nothing.  
\begin{equation}
\label{eq:cas}
	%\CASxre = \langle \If x = r \Then (x \asgn e \scomp \fence) \rangle
	\CASxre \sdef
	\atomic{\guard{x = r} \atomicsep x \asgn e \atomicsep \fence}
	\choice 
	\guard{x \neq r} 
\end{equation}
When used as the expression in a conditional we use the following abbreviation.
\begin{equation}
\begin{split}
\label{eq:tso-cas}
	~& \If \CAS{x}{ r}{ e} \Then c_1 \Else c_2
	\sdef
	\\ 
	~& 
	\qquad\qquad\qquad
	(\atomic{\guard{x = r} \atomicsep x \asgn e \atomicsep \fence} \cbef c_1 )
	\choice 
	(\guard{x \neq r} \cbef c_2 )
\end{split}
\end{equation}

\section{Operational semantics}
\label{semantics}

\begin{figure}[thp]
\begin{centering}

\newcommand{\RuleSep}{\vskip 2.5mm}

\ruledefNamed{110mm}{Prefix with reordering}
{reorder-rule-2}{
	\prefixac \tra{\aca} \cmdc
	~~~~~~~(a)
	\qquad
	\Rule{
		\cmdc \tra{\acb} \cmdc'
		\quad
		\aca \ro \fwd{\aca}{\acb}
	}{
		\prefixac \ttra{\fwd{\aca}{\acb}} \prefixacp
	}
	%\Rule{
		%\cmdc \tra{\acb} \cmdc'
		%\quad
		%\aca \ro \acb
	%}{
		%\prefixac \ttra{\fwd{\aca}{\acb}} \prefixacp
	%}
	~~~~~~~(b)
}

%\end{equation}
\ruledefNamed{40mm}{Choice}
{nondet}{
	\begin{array}{lcl}
	%\begin{array}{lclcclcl}
	\cmdc \choice \cmdd &\tra{\tau}& \cmdc
	\\
	%&\quad&
	\cmdc \choice \cmdd &\tra\tau{}& \cmdd
	\end{array}
}
\ruledefNamed{88mm}{Loop}
{loop-unfold-rule}{
	\begin{array}{rl}
	&
	\WHbc 
	\\
	\tra{\tau} 
	&
	(\guard{b} \cbef \cmdc \scomp \,\, \WHbc) \choice (\guard{\neg b} \cbef \Skip)
	\end{array}
}

\ruledefNamed{0.48\textwidth}{Locals - update}
{locals-reg}{
	\Rule{
		\cmdc \ttra{r \asgnlbl v} \cmdc'
	}{
		\locals{\sigma}{\cmdc}
		\tra{\tau}
		\locals{\Update{\sigma}{r}{v}}{\cmdc'}
	}
}
\ruledefNamed{0.48\textwidth}{Locals - store}
{locals-store}{
	\Rule{
		\cmdc \ttra{x \asgnlbl r} \cmdc'
		\quad
		\sigma(r) = v
	}{
		\locals{\sigma}{\cmdc}
		\ttra{x \asgnlbl v}
		\locals{\sigma}{\cmdc'}
	}
}

\ruledefNamed{0.50\textwidth}{Locals - load}
{locals-load}{
	\Rule{
		\cmdc \ttra{r \asgnlbl x} \cmdc'
	}{
		\locals{\sigma}{\cmdc}
		\ttra{\Readxv}
		\locals{\Update{\sigma}{r}{v}}{\cmdc'}
	}
}
\ruledefNamed{0.48\textwidth}{Locals - guard}
{locals-guard}{
	\Rule{
		\cmdc \ttra{\guardlble} \cmdc'
		%\quad
		%e_\sigma \nequiv false
	}{
		\locals{\sigma}{\cmdc}
		\ttra{\guardlbl{e_{\sigma}}}
		\locals{\sigma}{\cmdc'}
	}
}

%\RuleSep

%\end{equation}
\ruledefNamed{0.35\textwidth}{Thread id}
{threads}{
	\Rule{
		p \tra{\aca} p'
	}{
		\thread{\pidn}{p}
		\ttra{\pidn:\aca}
		\thread{\pidn}{p'}
	}
}
\ruledefNamed{0.62\textwidth}{Interleave parallel}
{pl}{
	\Rule{
		\prp_1 \tra{\aca} \prp_1'
	}{
		\prp_1 \pl \prp_2 \tra{\aca} \prp_1' \pl \prp_2
	}
	\quad
	\Rule{
		\prp_2 \tra{\aca} \prp_2'
	}{
		\prp_1 \pl \prp_2 \tra{\aca} \prp_1 \pl \prp_2'
	}
}

\end{centering}
\caption{Semantics of the language}
\label{fig:semantics-main}
\end{figure}

The meaning of our language is formalised using an operational semantics, which, excluding the global storage system, is
summarised in \reffig{semantics-main}.
Given a program $\cmdc$ the operational semantics generates a \emph{trace}, \ie, a possibly infinite sequence of steps
$\cmdc_0 \tra{\aca_1} \cmdc_1 \tra{\aca_2} \ldots$ where the labels in the trace are actions, or  
a special label
$\tau$ representing a silent or internal step that has no observable effect.  
For brevity we omit rules that are a straightforward promotion of a label from a subterm to a parent term, i.e., rules of the form 
$p \tra{\lbl} p' \imp C[p] \tra{\lbl} C[p']$.
%\subsubsection{Thread-local behaviour}

The terminated command $\Skip$ has no behaviour; a trace that ends with this command is assumed to have completed.
The effect of instruction prefixing in \refrule{reorder-rule-2} is discussed in \refsect{overview-rule}.
Note that actions become part of the trace.  
%We describe an instantiation for reordering and forwarding corresponding to the semantics
%of ARM and POWER in \refsect{reordering-forwarding}.

A nondeterministic choice (the \emph{internal choice} of CSP \cite{CSP}) can choose either branch, as given by \refrule{nondet}.
The semantics of loops is given by unfolding, e.g., \refrule{loop-unfold-rule} for a `while' loop.
Note that {\em speculative execution\/} is theoretically unbounded, and loads from inside later iterations of the loop could occur in earlier
iterations.  

%\subsubsection{Thread-level behaviour}

%Operations on local variables can be calculated within a process, using the syntax of the labels to remove the local and global state from the
%configurations.  
%This is similar to an idea used by Owens \cite{OwensOcamllight} and Abadi \& Harris \cite{AbadiHarris2009} 
%to simplify operational semantics.
For ease of presentation in defining the semantics for local states, we give rules for specific forms of actions, i.e.,
assuming that $r$ is a local variable in the domain of $\sigma$, and that $x$ is a global (not
in the domain of $\sigma$).
The more general version
can be straightforwardly constructed from the principles below.  

\refrule{locals-reg}
states that an action updating variable $r$ to value $v$ results in a change to the local state (denoted $\Update{\sigma}{r}{v}$).  Since this is a purely local operation there
is no interaction with the storage subsystem and hence the transition is promoted as a silent step $\tau$.
\refrule{locals-store} states that a \emph{store} of the value in variable $r$ to global $x$ is promoted as an instruction $x \asgn v$
where $v$ is the local value for $r$.
\refrule{locals-load} covers the case of a 
\emph{load} of $x$ into $r$.  The value of $x$ is not known locally.  
The promoted label is a
guard requiring that
the value read for $x$ is $v$.  This transition is possible for any value of $v$, but the correct value will be resolved when the label is
promoted to the storage level.
\refrule{locals-guard} states that a guard is partially evaluated with respect to the local state before it is promoted to the global level.  The notation $\evalse$ replaces $x$ with $v$ in $e$ for all $(x \mapsto v) \in \sigma$.

\refrule{threads} simply tags the process id to an instruction, to assist in the interaction with the storage system, and otherwise has no effect. Instructions of concurrent processes are interleaved in the usual way as described by \refrule{pl}.  

Other straightforward rules which we have omitted above include the promotion of
fences through a local state, and that atomic sequences of actions are handled inductively by the above rules. 

\subsection{Multi-copy atomic storage subsystem.}
\label{storage-subsystem-semantics}

\begin{figure}[t]
\begin{centering}

\ruledefNamed{85mm}{Globals - store}
{globals-store}{
	\Rule{
		p \ttra{\pidn: x \asgnlbl e} p'
	}{
		\globals{\sigma}{p}
		\tra{\tau}
		\globals{\Update{\sigma}{x}{\evalse}}{p'}
	}
}

\ruledefNamed{72mm}{Globals - guard}
{globals-guard}{
	\Rule{
		p \ttra{\pidn: \guardlble} p'
		\quad
		\evalse  \equiv true
	}{
		\globals{\sigma}{p}
		\tra{\tau}
		\globals{\sigma}{p'}
	}
}

\end{centering}
\caption{Semantics of a standard (multicopy-atomic) storage system}
\label{fig:semantics-multicopy-atomic}
\end{figure}

Traditionally, changes to shared variables occur on a shared global state, and when written to the global state are seen instantaneously by all processes in the system.  This is referred to as {\em multi-copy atomicity\/} and is a feature of TSO and the most recent version of ARM \cite{ARMv8.4}.
Older versions of ARM and
POWER, however, lack such multi-copy atomicity and require a more complex semantics.  We give the simpler case (covered in
\reffig{semantics-multicopy-atomic}) 
first.
For the store model the thread ids are not used, but they do become important in later sections.

Recall that at the global level the process id $\pidn$ has been tagged to the actions by \refrule{threads}.  
\refrule{globals-store} covers a store of some expression $e$ to $x$.  
Since all local variable references have been replaced by their values at the process level due to Rules~(\ref{rule:locals-reg})-(\ref{rule:locals-guard}),
expression $e$ must refer only to shared variables in $\sigma$.  The value of $x$ is updated to the fully evaluated value, $\evalse$.  

\refrule{globals-guard} states that a guard transition $\guarde$ is possible exactly when $e$ evaluates to true in
the global state.  If it does not, no transition is possible; this is how incorrect branches are eliminated from the
traces, which we discuss in more detail in the context of \emph{speculative execution} in \refsect{speculative-execution}.

\subsection{Reordering and forwarding (for sequential consistency)}
\label{reordering-forwarding}

\begin{figure}[t]

\begin{eqnarray}
	\fwd{y \asgnlbl f}{x \asgn e} &=& x \asgn \Repl{e}{y}{f}
		\quad
		%\mbox{~if}
		%\\ &&
		%\hspace{6pt}
		\mbox{if $e$ has no shared variables}
	\label{eq:fwd-u}
	\\
	\fwd{y \asgnlbl f}{\guarde} &=& \guard{\Repl{e}{y}{f}}
		\quad
		%\mbox{~if}
		%\\ &&
		%\hspace{6pt}
		\mbox{if $e$ has no shared variables}
	\label{eq:fwd-g}
	\\
	\notag
	\fwd{\aca}{ \acb} &=& \acb
	~~
		\mbox{otherwise}
\end{eqnarray}

\caption{Forwarding (bypassing)}
\label{fig:forwarding}
\end{figure}

It remains to define the reordering relation $\ro$ for particular architectures and the effect of forwarding so
that the effect of \refrule{reorder-rule-2} can be determined; and to model the global storage system where the
system lacks multi-copy atomicity defined by the rules given above.  We define the reordering relation in the following sections, though we may straightforwardly
define the reordering for an atomic sequence of instructions recursively as below, where $s$ is a sequence of instructions.
\begin{eqnarray}
	\aca &\ro& \eseq
	\\
	\aca \ro \seqT{\acb} \scat s
	&\iff&
	\aca \ro \acb \land \aca \ro s
\end{eqnarray}
We define similarly for the cases $s \ro \alpha$.

We note the trivial case for defining reordering for sequentially consistent (SC) processors: $\aca \nro
\acb$ for all $\aca, \acb$, and there is no forwarding.  Since reordering is not possible the second case of \refrule{reorder-rule} never
applies and hence the standard prefixing semantics is maintained.
SC semantics uses a storage system defined by the rules in \reffig{semantics-multicopy-atomic}.

Forwarding, as given in \reffig{forwarding}, is regular across all the architectures we have considered: $\fwdab$, where $\aca$ is an assignment $y \asgn f$
where $f$ does not contain shared variables, is straightforward replacement of $y$ by $f$ in the expression of an assignment \refeqn{fwd-u} or guard
\refeqn{fwd-g}.  Otherwise forwarding has no effect.  If forwarding was applied when $f$ contained shared variables, e.g., when $f$ is the expression $z$,
this would create more loads of $z$ resulting in potentially different values.  This approach to modelling forwarding contrasts with
an explicit FIFO buffer which is often used in modelling TSO.  The most recent store to a global $x$ is recorded in the program text, and need not be
explicitly kept separately in a buffer structure.

\subsection{Tool support and validation}
\label{validation-overview}

The operational semantics have been encoded in Maude \citep{Maude,SOSMaude} as rewrite rules.  A process in the language is rewritten to a trace, with the
Maude system generating all possible traces through backtracking.  To validate the semantics of particular architectures and to verify data structures running on them, we devised a straightforward mechanism for checking the final state against a
condition.

Modelling a particular architecture requires instantiating the reordering relation.  For commercial reasons, formal definitions of the hardware are not provided by the vendors. To establish confidence in our semantics, therefore,  we validated it
against litmus tests, small assembler programs.  We are fortunate in that considerable effort has gone into testing real hardware, collecting the results,
and using these to fine-tune an understanding of the hardware for TSO, ARM and POWER.  
However, it must be noted that the use of litmus tests, and their results on hardware, are problematic for validation for several reasons:
\begin{itemize}
\item
The set of litmus tests is unlikely to be complete.
\item
There may be a bug in the particular hardware tested giving incorrect behaviour.
\item
The particular hardware tested may not implement all features allowed by the memory model.
\item
The ``specification'' of the hardware may have imprecisions that resulted in vendors allowing behaviours that were intended to be forbidden.
%The ``specification'' of the hardware may incorrectly allow certain behaviours.
\item
Expected behaviours can change as new hardware is released.
\item
The absence of a behaviour does not mean that it is forbidden.
\end{itemize}
%Nevertheless, the sets we use are taken from rigorous and systematic efforts by other researchers.

Due to the above limitations, we do not attempt to achieve full conformance to the hardware results reported (where the above limitations
are also noted), nor do we try to match exactly the results of other models -- indeed some of the models themselves do not achieve full conformance, in
particular allowing many behaviours that were not observed on hardware.  Instead we aim to agree with litmus tests in the majority of cases, noting that 
refining a model to agree on all known litmus tests may quickly become redundant due to reasons above.

The litmus tests are provided in assembler syntax which we must translate to our wide-spectrum language. Branch instructions (e.g., ARM's \T{BNE}) are modelled using a combination of guards and nondeterministic choice.
A guard $\guarde$, where $e$ is an expression, does not directly map from a hardware
instruction.  Abstractly, a command $(\guard{r = 0} \cbef c)$ means that if $r = 0$ in the local state then $c$ may continue execution.  If $r \neq 0$, then no
execution is possible.  As such, our guard corresponds to the guards in Dijkstra's guarded command language \citep{GuardedCommands}.  
In our language we can use guards to model branching provided a straightforward structure is used, as outlined below.
Let $\aca_i$ stand for instructions.  The \T{BNE L} instruction jumps to label \T{L} if a special register is not equal to 0, while \T{B L}
unconditionally jumps to \T{L}. Thus a structure such as the following
\begin{align}
	\aca_1 \cbef \T{BNE~L1} \cbef \aca_2 \cbef \T{B~L2} \cbef \T{L1:~} \aca_3 \cbef \T{L2:} \ldots
%\end{equation}
\\
\quad
\mbox{
becomes 
}
\quad
%\begin{equation}
	\aca_1 \cbef (\If cmpr = 0 \Then \aca_2 \Else \aca_3) \cbef \ldots 
\end{align}
We have used the name $cmpr$ for the local register implicitly accessed by the \T{BNE} instruction.
Note that in our framework a branch instruction such as \T{BNE} in the assembler code structurally corresponds to a
guard (covering the true and false cases). 
This structured programming approach to denoting branching cannot cover all possible jumps within hardware addresses, but is sufficiently expressive to
capture the behaviours found in the litmus tests, and is suitable for modelling higher-level structured code.  

%All litmus tests were run on a Intel Xeon E5-2760 v3 with 384GB RAM.  

\section{TSO}
\label{tso}

\begin{figure}[th]
\noindent

%\begin{minipage}{0.30\linewidth}
\begin{eqnarray}
	\aca &\nro& \fence 
	\label{eq:a<f-tso}
	\\
	\fence &\nro& \aca
	\label{eq:f<a-tso}
	\\
	\guard{b} &\nro& \aca
	\label{eq:g<a-tso}
	\\
	\aca &\nro& \guardb 
	\label{eq:a<g-tso}
	\\
	%\aca &\ro& \acb 
		%~~
		%\mbox{in all other cases}
		%\notag
%\end{eqnarray}
%\end{minipage}
%\break
%\begin{minipage}{0.67\linewidth}
%\begin{eqnarray}
	x \asgn e &\ro& r \asgn f 
		%\mbox{~iff}
		\quad
	\label{eq:u<u-tso}
		%\\ &&
		%\hspace{14mm}
		\mbox{if $\vnfixf$, $\vnfi{r}{e}$, and}
		\\&&
		\hspace{14mm}
		\mbox{$e$ has no free globals}
		\notag
		%\\&&
		%\hspace{4pt}
		%\mbox{if $e$ has shared variables}
		%\notag
		%\\&&
		%\hspace{8pt}
		%\mbox{then $x \nfi f$}
		%\notag
	\\
	\aca &\nro& \acb
	\quad
	\mbox{in all other cases}
\OMIT{
	\\
	&& \notag \\
	\fwd{x \asgnlbl e}{r \asgn f} &=& f \asgn \Repl{f}{x}{e}
		\mbox{~if}
	\label{eq:fwd-u-tso}
		\\ &&
		\hspace{6pt}
		\mbox{$e$ has no shared variables}
		\notag
	\\
	\notag
	\fwd{\aca}{ \acb} &=& \acb
	~~
		\mbox{otherwise}
}
\end{eqnarray}
%\end{multicols}
%\end{minipage}
\caption{
Reordering following TSO assembler semantics.
$x$ denotes any variable, and $r$ a local variable.
}
\label{fig:tso-ro}
\end{figure}

The reordering relation for TSO is given in \reffig{tso-ro}. 
%In TSO the basic instructions are stores and loads, which are simple assignment statements where there are no globals in the stored expression and the
%loads are into local registers, with exactly one global in the load expression.  That is, we define
%\[
%	\tsostorexv \sdef x \asgn v
%	\qquad
%	\tsoloadry \sdef r \asgn y
%	\qquad
%	\tsofence \sdef \fence
%\]
%and similarly for register operations.  As TSO does not have speculative execution it does not require control fences.  
It uses a multi-copy atomic storage
system as defined by the rules in \reffig{semantics-multicopy-atomic}.
TSO is a relatively strong memory model with $\aca \nro \acb$ for all $\aca,\acb$ except as specified in \refeqn{u<u-tso}, which
allows loads to come before independent stores.

In addition \refeqn{u<u-tso} allows independent register operations to also be reordered before stores, allowing forwarding (or bypassing).
This means that a load of $x$ may take the value of the most recently written value to $x$. In our framework, this means that
if a load of $x$ is reordered before a store to $x$,
it takes that value.  
That is, since
\begin{equation}
	\fwd{x \asgn 1}{r \asgn x} = r \asgn 1
\end{equation}
from \refeqn{fwd-u}, we have
\begin{equation}
	x \asgn 1 \cbef r \asgn x
	\refsto
	r \asgn 1 \bef x \asgn 1
\end{equation}
by \reflaw{swap-order}.
Note that the instruction type changes from a load ($r \asgn x$) to a simple update to a local register ($r \asgn 1$), and hence is not affected by any
earlier stores to $x$.  

TSO's $\fence$ instruction can be employed to prevent the reordering of stores and loads \refeqns{a<f-tso}{f<a-tso}. 

\subsection{Validation}

We tested our definitions for TSO against the litmus tests mentioned in \cite{x86-TSO} and 25 generated tests using the \T{herd} tool (\url{http://diy.inria.fr/herd/}).
Those litmus tests cover the essence of TSO, namely that loads can appear to come before stores, and forwarding (or bypassing) takes place.

\section{Revised ARM v8}
\label{armv8}

%The multiprocessor architecture ARM is very relaxed in the orderings it allows.  In addition its storage system lacks \emph{multi-copy atomicity}, 
%meaning that different
%threads may see updates to variables in different orders.  We consider these two aspects separately.

In this section we consider the latest (revised) version of ARM v8 which is multi-copy atomic \cite{ARMv8.4}. We consider older versions of ARM which lack multi-copy atomicity in \refsect{arm}.

In addition to stores, loads, register operations, and full fences, ARM's instruction set includes a \emph{control fence}, $\cfence$,
which affects local reordering by acting as a barrier 
preventing subsequent loads being reordered with earlier instructions.  It is used in conjunction with branches to avoid the effect of
speculative execution, discussed in \refsect{speculative-execution}.  ARM also has a \emph{store only} fence.

\begin{figure}[thp]
%\begin{minipage}{0.45\linewidth}
\begin{eqnarray}
	\aca &\ttdef& \ldots \csep \wwfence \csep \cfence
	\label{eq:cfence-syntax}
	\label{eq:wwfence-syntax}
	\\
	\Also
	\aca &\nro& \fence 
	\label{eq:a<f}
	\\
	\fence &\nro& \aca
	\label{eq:f<a}
	\\
	x \asgn e &\nro& \wwfence 
		\ltif{$x$ is shared}
	\label{eq:a<wwf}
	\\
	\wwfence &\nro& x \asgn e
		\ltif{$x$ is shared}
	\label{eq:wwf<a}
	\\
	\guardb &\nro& \cfence
	\label{eq:g<cf}
	\\
	\cfence &\nro& r \asgn e
	\label{eq:cf<l}
	\\
	\guard{b_1} &\ro& \guard{b_2}
		\ltif{$\loadDistinct{b_1}{b_2}$}
	\label{eq:g<g}
	\\
	%\guardb &\nro& \gvar \asgn e
	\guardb &\nro& x \asgn e
		\ltif{$x$ is shared}
	\label{eq:g<s}
	\\
	\guardb &\ro& r \asgn e
		\ltif{$\vnfirb$ and $\loadDistinct{e}{b}$}
	\label{eq:g<r}
	\\
	x \asgn e &\ro& \guardb
		\ltif{$\vnfix{b}$ and $\loadDistinct{e}{b}$}
	\label{eq:u<g}
	\\
%\end{eqnarray}
%\end{minipage}
%\break
%\begin{minipage}{0.53\linewidth}
%\begin{eqnarray}
	x \asgn e &\ro& y \asgn f 
		\mbox{~if}
	\label{eq:u<u}
		%\\ &&
		\hspace{6pt}
		\mbox{$\varDistinct{x}{y}$, $\vnfixf$, $\vnfiye$, and}
		%\notag
		%\\&&
		%\hspace{6pt}
		\mbox{~$\loadDistinctef$}
		%\notag
		%\\&&
		%\hspace{4pt}
		%\mbox{if $e$ has shared variables}
		%\notag
		%\\&&
		%\hspace{8pt}
		%\mbox{then $x \nfi f$}
		%\notag
	\\
	\aca &\ro& \acb 
		~~
		\mbox{in all other cases}
		\notag
	\OMIT{
	&& \notag \\
	\fwd{y \asgnlbl f}{x \asgn e} &=& x \asgn \Repl{e}{y}{f}
		\mbox{~if}
	\label{eq:fwd-u}
		\\ &&
		\hspace{6pt}
		\mbox{$e$ has no shared variables}
		\notag
	\\
	\fwd{y \asgnlbl f}{\guarde} &=& \guard{\Repl{e}{y}{f}}
		\mbox{~if}
	\label{eq:fwd-g}
		\\ &&
		\hspace{6pt}
		\mbox{$e$ has no shared variables}
		\notag
	\\
	\notag
	\fwd{\aca}{ \acb} &=& \acb
	~~
		\mbox{otherwise}
	}
\end{eqnarray}
%\end{multicols}
%\end{minipage}
\caption{
Reordering following ARM assembler semantics. $x,y$ denote any variable and $r$ a local variable.
}
\label{fig:higher-level-defs}
\end{figure}

Our general semantics is instantiated for ARM processors in \reffig{higher-level-defs} which provides particular definitions for the reordering relation that are generalised from the orderings on stores and loads in these processors. 
%This reordering relation is identical to that of POWER processors considered in \refsect{power}.%
%\footnote{We have excluded address shifting, which creates \emph{address dependencies} \cite{HerdingCats}, as 
%this does not affect the majority of high-level algorithms in which we are interested. 
%However, address dependencies are accounted for in our tool as discussed in \refsect{address-dependencies}.}

Fences prevent all reorderings as with TSO \refeqns{a<f}{f<a}, while a store-only barriers $\wwfence$ (corresponding to 
ARM's \T{DMB.ST} and \T{DSB.ST} instructions) maintains order on stores but not on other instruction types \refeqns{a<wwf}{wwf<a}. 
A control fence $\cfence$ prevents speculative loads when placed between a guard and a load \refeqns{g<cf}{cf<l}.  
Guards may be reordered with other guards provided they do not both access the same shared variables \refeqn{g<g} (otherwise local coherence would be violated), 
but stores to shared variables may not come before a guard evaluation \refeqn{g<s}.
This prevents speculative execution from modifying the global state, in the event that the speculation was down the wrong branch.
An update of a local variable may be reordered before a guard provided it does not affect the guard expression and respects local coherence \refeqn{g<r}.
Guards may be reordered before updates if those updates do not affect the guard expression and local coherence is respected \refeqn{u<g}.
(Note that in ARM assembler the $\loadDistinct{e}{b}$ constraints for guards are always satisfied as guards (branch points) do not reference globals.)
Assignments may be reordered as shown in \refeqn{u<u} and discussed in \refsect{overview-reordering}.

\subsection{Speculative execution}
\label{speculative-execution}

Many processors allow some form of \emph{speculative execution}, where the instructions in a branch are
tentatively executed and the effect stored locally while, for instance, waiting for a load of a global to be
serviced.  On TSO and related architectures the result of speculative execution are not visible, i.e.,
speculatively executed loads are restarted if it is detected that an old value was loaded.  On ARM processors the effect of speculative execution can become visible, i.e., the
effect of speculatively executing loads is not (conditionally) unwound.  However, in all cases, if speculation
was down a branch that was eventually determined to be incorrectly chosen, no effect is (or should be) visible.\footnote{The recently discovered Spectre security vulnerability \cite{SpectreCoRR} shows that this is not strictly the case.}
Fortunately this complication can be handled straightforwardly in our semantics.

If a guard does not evaluate to $true$, execution stops in the sense that no
transition is possible.  This corresponds to a false guard, i.e., $\Magic$ \cite{Morgan:94,Back:98}, and such
behaviours do not terminate and are ignored for the purposes of determining behaviour of a system.
Interestingly, this simple concept from standard refinement theory allows us to handle speculative execution
straightforwardly.  In existing approaches, the semantics is complicated by needing to restart reads if speculation
proceeds down the wrong path.  Treating branch points as guards works because speculation should have no effect if the
wrong branch was chosen.  

\label{load-speculation}
\newcommand{\Mark}[1]{\underline{#1}}

To understand how this approach to speculative execution works, consider the following deri\-vation.  
%Assume that (a) loads may be reordered before guards if they reference independent variables, and (b) loads may be reordered if they reference different variables.  Recall that we omit trailing $\Skip$ commands to save space.
\begin{derivation}
	\step{
		r_1 \asgn x \cbef (\If r_1 = 0 \Then \Mark{r_2 \asgn y})
	}
	\trans{=}{Definition of $\If$ \refeqn{defn-if}}
	\step{
		r_1 \asgn x \cbef ((\guard{r_1 = 0} \cbef \Mark{r_2 \asgn y}) \choice \guard{r_1 \neq 0})
	}
	\trans{\refsto}{Resolve to the first branch, since $(\cmdc \choice \cmdd) \refsto \cmdc$ \OMIT{by \refeqn{defn-reft}}}
	\step{
		r_1 \asgn x \cbef \guard{r_1 = 0} \cbef \Mark{r_2 \asgn y} 
	}
	\trans{\refsto}{From \reflaw{swap-order} and (52)} %\refeqn{g<l}}
	\step{
		r_1 \asgn x \cbef \Mark{r_2 \asgn y} \bef \guard{r_1 = 0} 
	}
	\trans{\refsto}{From \reflaw{swap-order} and (54)}%\refeqn{l<l}; \reflaw{keep-order}}
	%\footnote{This step requires a specialisation of \reflaw{swap-order} which has $(\alpha;\beta \bef c)$, rather than $(\alpha;\beta;c)$, as its left-hand side.} 
	\step{
		\Mark{r_2 \asgn y} \bef r_1 \asgn x \cbef \guard{r_1 = 0} 
	}
	%\trans{\refsto}{From \reflaw{keep-order}}
	%\step{
		%\Mark{r_2 \asgn y} \bef r_1 \asgn x \bef \guard{r_1 = 0} \bef \cfence
	%}
\end{derivation}
This shows that the inner load (underlined) may be reordered before the branch point, and subsequently before an earlier load.
Note that this behaviour results in a terminating trace only if $r_1 = 0$ holds when the guard is evaluated, and otherwise becomes $\Magic$ (speculation down
an incorrect path).
On ARM processors, placing a control fence ($\cfence$) instruction inside the branch, before the inner load, prevents this reordering.

\subsection{Address shifting} 
\label{address-shifting} 
\label{address-dependencies}

In ARM the address an instruction loads from (or stores to) may be shifted.
%For the majority of high-level algorithms such details are hidden.  However address shifting is investigated at the hardware level because it can affect reordering --
%so called ``address dependencies'' \citep{HerdingCats}.  
The instruction \T{LDR~R1,~[R2,~X]} loads into \T{R1} the value at address \T{X} shifted by the amount in \T{R2}.  
The presence of address shifting, or other mechanisms for modifying the address target of an instruction, can have an influence on the reordering relation, as captured by the
\T{addr} relation in \cite{HerdingCats}.  In our framework most of the restrictions introduced by address shifting are already captured by \refrule{reorder-rule-2}, where we interpret
`$x$' to range over expressions including address shifts such as $\addrShift{r}{y}$, and hence the set of variables being checked against the conditions is $\{y,r\}$.
However, as mentioned in \cite{ModellingARMv8}, an instruction $\addrShift{e}{x} \asgn f$ has at least one other effect on reordering, which is that later stores cannot be
reordered before it, in case the shift amount $e$ gives an invalid address and results in an exception being thrown.  In such cases, the effect of later writes should not be
visible to other processes.
\[
	\addrShift{e}{x} \asgn f \nro y \asgn g
		\quad \mbox{if $y$ is a shared variable}
\]
To precisely model the semantics of address shifting requires a more concrete model than the one we propose, however, as determined by the litmus tests of
\citep{ModellingARMv8}, the effects of address shifting on reorderings can be investigated even when the shift amount is 0 (resulting in a load of the value at the address).
As such we define $\addrShift{0}{x} = x$, and leave the effect of other shift amounts undefined.  
Note that when the shift amount expression is evaluated to 0 (e.g., by forwarding) the address shifting is removed and this can have an effect on the allowed reorderings.
%As such we define that an address shift of 0 on a variable $x$ gives $x$, and leave the effect of other shift amounts undefined.  

%Address shifting constrains the reorderings in the following ways: a branch
%may not be reordered before a load or store with address shifting; a store (to any variable) may not be reordered before an
%instruction with address shifting; and any instruction $\aca$ which shares a register or variable with $\acb$ where $\acb$ has
%address shifting may not be reordered with $\acb$.  We incorporate these conditions into the general rules for reordering assignments and branches.

%A further consequence of address shifting is that a load $\load{r_2}{x}$ may be reordered before $\load{r_1}{\addrShift{n}{x}}$ even though this would 

\subsubsection{Load speculation}
\label{load-load-semantics}

A further aspect of address shifting is that in some circumstances a load $r_2 \asgn x$ may be reordered before a load $r_1 \asgn \addrShift{e}{x}$,
even though this would appear to
violate coherence-per-location.  However, the load into $r_2$ must not load a value of $x$ that was written before the value read
by the load into $r_1$.  This complex situation is handled in \citep{ModellingARMv8} by restarting load instructions if an earlier value is read into $r_2$.  We handle it more abstractly by
treating the load as speculation, where if an earlier value for $r_2$ is loaded then the effect of that speculation is thrown away (the point of
allowing this reordering is apparently to allow execution after the second load to continue while the value of the shift amount $e$ is
calculated).  
This is given by the following operational rule.
\begin{equation}
	\Rule{
		\prp \ttra{\loadlbl{r_2}{x}} \prp'
	}{
		r_1 \asgn \addrShift{n}{x} \cbef \prp
		\ttra{\loadlbl{r_2}{x}}
		r_1 \asgn \addrShift{n}{x} \cbef \guard{r_1 = r_2} \cbef \prp'
	}
\end{equation}
In practical terms it is possible the first load of $x$ (into $r_1$) is delayed while determining the offset value.
The later load is allowed to proceed, freeing up $\prp'$ to continue speculatively executing until the dependency is resolved.  The load
into $r_1$ then must still be issued, the result being checked against $r_2$.  This check must occur to preserve coherency as the load into $r_2$ cannot
read a value earlier than that read into $r_1$.  Note that loads in $\prp'$ can now potentially be reordered to execute ahead of the load into $r_1$.

\OMIT{
Algebraically:
\[
	\load{r_1}{x} \cbef (\load{r_2}{x} \bef \prp)
	= 
	\load{r_2}{x} \bef 
	(\load{r_1}{x} \cbef (\guard{r_1 = r_2} \cbef \prp'))
\]
	
}

\subsection{Eliminating earlier writes}
\label{elimEarlierWrites}

An additional aspect of ARM processors is that when there are consecutive writes to a variable $x$ on a process the first write can effectively be eliminated: locally only
the effect of the second write will be seen (sequential semantics is preserved), and globally it is always a valid behaviour that a sibling process did not see the effect of the first write because the second
occurred immediately after it.  Write elimination is captured by the following rule.
\begin{equation}
	\Rule{
		c \ttra{x \asgnsmall v} c'
	}{
		x \asgn w \cbef c
		\ttra{x \asgnsmall v}
		c'
	}
\end{equation}
We may derive the following elimination law.
\begin{equation}
	x \asgn w \cbef x \asgn v \cbef c ~~\refsto~~ x \asgn v \cbef c
\end{equation}

\subsection{Validation}
\label{arm-validation}

The new version of ARM as reported in \cite{ARMv8.4} is quite recent and the litmus tests used in that paper are not available at the time of writing.  We
have validated earlier versions of ARM which are more complicated, and as such we defer discussion of validation until \refsect{armv8-validation}.

\section{Original ARM v8 and earlier -- non-multicopy atomicity}
\label{arm}

In this section we consider the versions of ARM which lack multi-copy atomicity. These include the original version of ARM v8 \cite{ModellingARMv8} and all earlier versions. These versions of ARM allow processes to communicate values to each other without accessing the heap.  
That is, if process $\prp_1$ is storing $v$ to $x$, and process $\prp_2$ wants to
load $x$ into $r$, $\prp_2$ may preemptively load the value $v$ into $r$, before $\prp_1$'s store hits the global shared storage.  
Therefore different processes may have
different views of the values of global variables; see litmus tests such as the \T{WRC} family \cite{HerdingCats}.  
%In addition to thread-local reordering, there is also an extra level of reordering on stores, since processes do not see updates by a thread
%simultaneously (due to the lack of multi-copy atomicity).  
To properly model these versions of ARM we must therefore introduce the storage subsystem, 
$\storage{\wseq}{c}$, which replaces the $\globals{\sigma}{\prp}$ notation defined earlier.

\subsection{Storage subsystem}
\label{storage-subsystem}

We conceptualise the stores in the system as a list $\wseq$ of \emph{writes} $w_1, w_2, \ldots$, with each write $w_i$ being of the form
\begin{equation}
	\srqnxvns
\end{equation}
where $\pidn$ is the process id of the thread that executed a store of value $v$ to address $x$.  The list $\pidns$ is a list of process ids that have \emph{seen}
this write, that is, loaded that value into some local register.
For such a write $w$, we let $w.var=x$, $w.thread=\pidn$ and $w.seen=\pidns$.
For a write $\srqnxvns$ it is always the case that $\pidn \in \pidns$. 

%\rcnote{In the FM paper and here we have new writes being inserted from the "front" of $\wseq$ (from index 1).  However traces are constructed by placing new
%elements on the end; also in the Maude encoding writes are inserted from the ``back'', which is standard for concatenating onto a sequence.  Should we change
%it here, the main drawback being incompatibility with the FM version?  Also, the recursive definitions for $flush_n$ etc. would have to change to work on
%the last element
%rather than the first}

The order of writes in $\wseq$ and previous values seen by a process affect the values it loads, which in general are nondeterministic.
When a new write $w$ is executed by a thread, $w$ is not necessarily appended to the end of $\wseq$, but instead may be ``inserted'' earlier,
according to certain rules.  
The basic principles of inserting a new write $w$ of the form
$\srqnxvn$
into the list $\wseq$ are:
\begin{enumerate}
\item
Request $w$ may not come before any earlier write by $\pidn$ (local coherence).  
\item
Request $w$ may not come before a write $w'$ to $x$ by another process \emph{that has been seen by $\pidn$} (global coherence).
\end{enumerate}

When process $\pidn$ loads the value of location $x$ from $\wseq$ it may see either the most recent value of $x$ that $\pidn$ has already seen, or any that have
been added 
more recently in $\wseq$.  When $\pidn$ sees some write $w$ then $\pidn$ is added to the list of seen process ids in $w$.  
The shared state from the perspective of a given process is a particular view of this list. There is no single definitive shared state.  In addition, viewing a value in the list causes the list to be updated 
and this affects later views.

Initially $\wseq$ holds writes giving the initial values of the shared variables.  These initial values are assumed to have been seen by
every process.

\newcommand{\canSeePast}[3]{canSeePast_#1(#2,#3)}
\newcommand{\canSeePastn}[2]{\canSeePast{\pidn}{#1}{#2}}
\newcommand{\canSeePastnx}[1]{\canSeePastn{x}{#1}}
\newcommand{\canSeePastnxw}{\canSeePastnx{w}}

\begin{figure}[tph]

\begin{eqnarray}
	\syss &\ttdef& \ldots 
		\cbar
		\storage{\wseq}{\prp}
\end{eqnarray}

\ruledefNamed{0.99\textwidth}{Storage - load}
{storage-load}{
	\Rule{
		p \ttra{\pidn:\Readxv} p'
		\also
		\all w \in \ran(\wseq_2) @ \canSeePastnxw
		%%\all w \in \ran(\wseq_1) @ x = w.var \imp \pidn \notin w.seen 
	}{
		\begin{array}{l}
		\storage{\wseq_1 \scat \srqmxv{\pidns} \scat \wseq_2}{p}
		\ttra{\pidn:\Readxv}
		\also
		%\qquad \qquad \qquad \qquad \qquad \qquad
		\storage{\wseq_1 \scat \srqmxv{\pidns \union \{\pidn\}} \scat \wseq_2}{p'}
		%\storage{\wseq_1 \scat w \scat \wseq_2}{p}
		%\ttra{\pidn:\Readxv}
		%\storage{\wseq_1 \scat w' \scat \wseq_2}{p'}
		\end{array}
	}
}

\ruledefNamed{0.99\textwidth}{Storage - store}
{storage-store}{
	\Rule{
		p \ttra{\pidn: x \asgnlbl v} p'
		\\
		w' = \srqnxvn
		\qquad
		\all w \in \ran(\wseq_2) @ \canReorder{w}{w'}
		%\all w \in \ran(\wseq_1) @ \canReorder{w}{\srqnxvn}
		%% \all w \in \ran(\wseq_1) @ 
			%% \pidn \neq w.thread 
			%% \land 
			%% (x = w.var \imp \pidn \notin w.seen)
	}{
		\storage{\wseq_1 \scat \wseq_2}{p}
		\ttra{\pidn: x \asgnlbl v}
		\storage{\wseq_1 \scat w' \scat \wseq_2}{p'}
		%\storage{\wseq_1 \scat \srqnxvn \scat \wseq_2}{p'}
	}
}

\ruledefNamed{0.99\textwidth}{Storage - fence}
{fence}{
	\Rule{
		p \ttra{\pidn:\fencelbl} p'
	}{
		\storage{\wseq}{p}
		\ttra{\pidn:\fencelbl}
		\storage{\flush_\pidn(\wseq)}{p'}
	}
}

where
\begin{eqnarray}
	\canSeePastnxw 
	&\sdef&
		x = w.var \imp \pidn \notin w.seen 
	\label{eq:def-canSeePast}
	\\
	\canReorder{w}{\srqnxvn} 
	&\sdef&
			\pidn \neq w.thread 
			\land 
			(x = w.var \imp \pidn \notin w.seen)
	\label{eq:def-wro}
	\\
	\flush_\pidn(\eseq) 
	= 
	\eseq
	\label{eq:base-flush}
	&&
	\flush_\pidn(\wseq \scat w) 
	= 
	\flush_\pidn(\wseq) \scat \flush_\pidn(w) 
	\\
	%\flush_\pidn(\srqmxvns) 
	\flush_\pidn(w) 
	&=&
		\left\{
		\begin{array}{ll}
			%\Update{w}{seen}{\PIDSet} \scat \flush_\pidn(\wseq) 
			%\srqmxv{\PIDSet}
			\Update{w}{seen}{\PIDSet} 
			&
			%if~ \pidn \in \pidns
			if~ \pidn \in w.seen
			\\
			%w \scat \flush_\pidn(\wseq) 
			%\srqmxvns
			w 
			& 
			otherwise
		\end{array}
		\right.
	\label{eq:def-flush}
	%\\
	%\flush_\pidn(\srqmxvns \scat \wseq) &=& \srqmxv{\PIDSet} \scat \flush_\pidn(\wseq) 
		%\quad if~ \pidn \in \pidns
	%\\
	%\notag
	%\flush_\pidn(\srqmxvns \scat \wseq) &=& \srqmxvns \scat \flush_\pidn(\wseq) 
		%\quad if~ \pidn \notin \pidns
\end{eqnarray}
\caption{Rules for the non-multi-copy atomic subsystem of ARM and POWER.}
\label{fig:storage-rules}
\end{figure}

We give two specialised rules (for a load and store) in \reffig{storage-rules}.
To handle the general 
case of an assignment $x \asgn e$, where $e$ may contain more than one shared variable, the 
antecedents of the rules are combined, retrieving the value of each variable referenced in $e$ individually and
accumulating the changes to $\wseq$.

\refrule{storage-load} states that process $\pidn$ can load the value $v$ for $x$ provided there is a write 
$\srqmxv{\pidns}$ 
in the system where all earlier writes can be ``seen past'', i.e., as given in \refeqn{def-canSeePast}, an earlier write to variable $x$ has not been seen.
As a result, $\pidn$ is added to the set of process ids that have seen that write (and hence later reads of $x$ by $\pidn$ will not be able to see any earlier writes
to $x$).

\refrule{storage-store} governs where a new write $w$ may be ``inserted'' into the global storage.
Write $w$ may appear earlier than writes that are already in system, provided it can be reordered with them as given in \refeqn{def-wro}.
We say $w$ can come before a write $v$, written $v \wro w$, provided $v$ was by another process, and if it is a write to the same variable as $w$ then it has not been
seen.
This constraint keeps all writes by a single process in the same order, and keeps a system-wide coherence on any one shared variable, but allows different processes
to see updates to different variables in a different order.
For instance, if the global storage contains writes to $x$ and $y$ by process $\pidn$,
\[
	\srqnxv{\pidns_1}
	\scat
	\srq{\pidn}{y}{w}{\pidns_2}
\]
then process $\pidm$ may see that $x$ has changed but read the initial value for $y$, while another process $\pid{p}$ may see that $y$ has changed but read the
initial value for $x$ (assuming neither $\pidm$ nor $\pid{p}$ are in $\pidns_1$ or $\pidns_2$).

\refrule{fence} states that
a $\fence$ action by process $\pidn$ `flushes' all previous writes seen by $\pidn$ (which includes those writes by $\pidn$). 
The $\flush$ function modifies $\wseq$ so that all processes can see all writes by $\pidn$, effectively overwriting earlier writes.
This is achieved by updating the write so that all processes have seen it, written as
$\Update{w}{seen}{\PIDSet}$, and defined recursively by \refeqns{base-flush}{def-flush}.
A $\wwfence$ instruction also flushes the storage as in \refrule{fence}.

\subsection{Validation}
\label{litmus-tests}
\label{armv8-validation}

The semantics was validated against 2 sets of litmus tests (with many overlapping tests).
The first was a set of 348 litmus tests developed in \citep{ModellingARMv8}
(we excluded some that could not
be automatically translated, and also exclude one that involves \emph{shadow registers}, for reasons described in \ref{shadow-registers}).
We compared our results (pass/fail) against the expected results on hardware taken from the supplemental material for \citep{ModellingARMv8}%
\footnote{
\url{https://www.cl.cam.ac.uk/~pes20/arm-supplemental/index.html}
}.
The translation process was 
straightforward, with conditional statements translated into guarded branches as described in \refsect{syntax-abbreviations}.  
In addition some redundant register
operations were eliminated to reduce tool time, for instance, $r \asgn 1 ; x \asgn r$ 
becomes $x \asgn 1$ provided $r$ is not used elsewhere in the code.

The majority of the tests (333) were performed in under 3s of processor time.  Those with 3 or more processes, limited local ordering (lack of
fences, etc.), and many writes, were the slowest. The Maude system measures rewrites and the largest test required approximately 50 million
rewrites, taking 33s.  Results of three of the 348 tests were not recorded in \citep{ModellingARMv8} as they did not complete in a
reasonable time.
%
%\footnote{
%These three tests (which are
%identical with each other in our model, since we do not distinguish \T{DMB} barriers from \T{DSB} barriers (both full fences), as in \citep{ModellingARMv8}) are not among the
%slower in our prototype tool.
%}
%We exclude from consideration tests involving ``shadow registers'', which appear to be processor-specific facilities which are not intended
%to conform to sequential semantics (they do not correspond to higher-level code): these are discussed in \ref{shadow-registers}.  
Of the remaining litmus tests all but three agreed with the model
results in \citep{ModellingARMv8}
which we discuss below in \refsect{ppo017}.

The second set of tests we used for validation was 
the set of 9790 tests from \cite{HerdingCats}.  These tests were used to validate an axiomatic semantics rather than an operational model as in \cite{ModellingARMv8}.
Our results for this larger set, excluding 5 that did not parse and 133 that did not complete in a reasonable amount of time, is 9556 tests are in agreement with
\cite{HerdingCats} and 96 in disagreement.  Of those 96, we agree with the results obtained by hardware in 52: 5 where we allow a behaviour seen on hardware (but
disallowed by \cite{HerdingCats}), the remainder where our model says such behaviours should be forbidden (and were not observed on hardware).  
Of the remaining 44, all but 3 involve a \T{store(x);store(x)} or \T{load(x);load(x)} pattern; of those 3, 2 are identical (\T{MP+dmb+addr-po-ctrlisb} and \T{MP0110}).  Those 2
identical tests are allowed in our model possibly erroneously, as they allow load speculation in the presence of an unresolved write with
address shifting.  
The remaining test, \T{MP+PPO015}, looks like it is disallowed by us erroneously because we do not have ``chain forwarding", or possibly
forwarding from a register assignment, which is necessary
\emph{inside a branch} to resolve an address shift expression.

If we instead exclude from the 96 disagreements those tests involving address shifting we are left with 20 tests where we disagree with \cite{HerdingCats}.  In all 20 cases
we allow the behaviour that is disallowed by \cite{HerdingCats} (and which has not been observed on hardware).  All of these 20 cases involve at least one instance of an access to a
shared variable twice or more in at least one process, i.e., \T{po-loc} becomes relevant.  By allowing more behaviours than \cite{HerdingCats}, we are erring on the side of soundness, i.e., 
if we can prove a program involving one of these 20 cases correct in our semantics, it is also certainly correct in the semantics of \cite{HerdingCats}.
%It is therefore hard to justify our behaviour against \HC.

We note that the results in \HC include 558 tests where their model forbids a result which was observed on hardware, and 1525 tests where the \HC model allows a behaviour which was
not observed. Regarding the former, the discrepancy is attributed to the
\emph{load-load hazard} (e.g., \cite{LLHazard}) 
and fewer discrepancies remain between their model and hardware when the affected litmus tests are excluded.   

\subsubsection{Discrepancies with \cite{ModellingARMv8}}
\label{ppo017}

\newcommand{\xor}{\mathrel{\mathtt{xor}}}

We present litmus test 
\T{PPO017} below,\footnote{
Available at
\url{
http://www.cl.cam.ac.uk/~pes20/arm-supplemental/src/PPO017.litmus
}} translated into our wide-spectrum language,
for which our model gives a different result to that of \citep{ModellingARMv8}.\footnote{
We simplified some of the syntax for clarity, in particular introducing a higher-level $\If$ statement to model a jump
command and implicit register (referenced by the compare (\T{CMP}) and branch-not-equal (\T{BNE}) instructions).
We have also combined some commands, retaining dependencies, in a way that is not possible in the assembler language.
The $\xor$ operator is exclusive-or; its use here (artificially) creates an \emph{address dependency} \cite{HerdingCats} between the updates to $r_0$ and $r_2$.
The $r_4 = r_4$ empty conditional creates a \emph{control dependency} (and with the control fence, a \emph{control fence dependency}) between loads.
}
This is structurally similar to the test \T{PPO015}, discussed in \cite{FM18}, and test \T{PPO012}: those two other tests pass in our model for essentially the same reason
as \T{PPO017}.
\begin{equation}
\begin{split}
\label{eq:1ppo017}
	~&
 x \asgn 1
 \cbef
 \fence
 \cbef
 y \asgn 1
 \quad
 \pl
 \\
 	~&
 r_0 \asgn y
 \cbef
 r_2 \asgn \addrShift{(r_0 \xor r_0)}{z}
 %z \asgn (r_0~{\tt xor}~r_0) + 1
 \cbef
 z \asgn 1
 \cbef
 r_4 \asgn z
 \cbef
  \\ 
  	~&
  \qquad \qquad
  (\If r_4 = r_4 \Then \Skip \Else \Skip)
  %(\guard{r3 = r3} \choice \guard{r_3 \neq r_3}) 
 \scomp
  \cfence
 \cbef
  r_5 \asgn x
\end{split}
\end{equation}
The tested condition is 
$r_0=1 \land r_5=0$, which asks whether it is possible to load $x$ (the last statement of process 2) before loading $y$ (the first statement of
process 2). At a first glance the control fence prevents the load of $x$ happening before the branch.  However, as indicated by litmus tests such as
\T{MP+dmb.sy+fri-rfi-ctrisb}, 
\cite{ModellingARMv8}[Sect 3,\emph{Out of order execution}], 
under some circumstances the branch condition can be
evaluated early.
We expand on this below by manipulating the second process, taking the case where the success branch of the $\If$ statement is chosen.
To aid clarity we underline the instruction that is the target of the (next) refinement step.
\begin{derivation*}
	\step{
 r_0 \asgn y
 \cbef
 %z \asgn (r_0~{\tt xor}~r_0) + 1
 r_2 \asgn \addrShift{(r_0 \xor r_0)}{z}
 \cbef
 z \asgn 1
 \cbef
 \underline{r_4 \asgn z}
 \cbef
  %}\step{
  %\qquad \qquad
  %(\If \T{r3 = r3} \Then \Skip \Else \Skip)
  \guard{r_4 = r_4} 
 \cbef
  \cfence
 \cbef
  r_5 \asgn x
	}

	\trans{\refsto}{Promote load with forwarding (from $z \asgn 1$), from \reflaws{keep-order}{swap-order}}

	\step{
 r_4 \asgn 1
 \bef
 r_0 \asgn y
 \cbef
 %z \asgn (r_0~{\tt xor}~r_0) + 1
 r_2 \asgn \addrShift{(r_0 \xor r_0)}{z}
 \cbef
 z \asgn 1
 \cbef
  %}\step{
  %\qquad \qquad
  %(\If \T{r3 = r3} \Then \Skip \Else \Skip)
  \underline{\guard{r_4 = r_4} }
 \cbef
  \cfence
 \cbef
  r_5 \asgn x
	}

	\trans{\refsto}{Promote guard by \reflaws{keep-order}{swap-order} (from \refeqn{u<g})}

	\step{
 r_4 \asgn 1
 \bef
  \guard{r_4 = r_4} 
 \bef
 r_0 \asgn y
 \cbef
 %z \asgn (r_0~{\tt xor}~r_0) + 1
 r_2 \asgn \addrShift{(r_0 \xor r_0)}{z}
 \cbef
 z \asgn 1
 \cbef
  %}\step{
  %\qquad \qquad
  %(\If \T{r3 = r3} \Then \Skip \Else \Skip)
  \underline{\cfence}
 \cbef
  r_5 \asgn x
	}

	\trans{\refsto}{Promote control fence by \reflaws{keep-order}{swap-order} (\refeqn{g<cf} does not now apply)}

	\step{
 r_4 \asgn 1
 \bef
  \guard{r_4 = r_4} 
 \bef
  \cfence
 \bef
 r_0 \asgn y
 \cbef
 %z \asgn (r_0~{\tt xor}~r_0) + 1
 r_2 \asgn \addrShift{(r_0 \xor r_0)}{z}
 \cbef
 z \asgn 1
 \cbef
  %}\step{
  %\qquad \qquad
  %(\If \T{r3 = r3} \Then \Skip \Else \Skip)
  \underline{r_5 \asgn x}
	}

	\trans{\refsto}{Promote load by \reflaws{keep-order}{swap-order} }

	\step{
 r_4 \asgn 1
 \bef
  \guard{r_4 = r_4} 
 \bef
  \cfence
 \bef
  r_5 \asgn x
 \bef
 r_0 \asgn y
 \cbef
 %z \asgn (r_0~{\tt xor}~r_0) + 1
 r_2 \asgn \addrShift{(r_0 \xor r_0)}{z}
 \cbef
 z \asgn 1
  %}\step{
  %\qquad \qquad
  %(\If \T{r3 = r3} \Then \Skip \Else \Skip)
	}

\end{derivation*}
The load $r_5 \asgn x$ has been reordered before the load $r_0 \asgn y$, and hence when interleaved with the first process from \refeqn{1ppo017}
it is straightforward that the condition may be satisfied.

In the Flowing/POP model of \cite{ModellingARMv8}, 
this behaviour is forbidden because there is an address dependency from the load of $y$ into \code{r_0} to \code{r_4}, via $z$.  
In the testing of real processors reported in \cite{ModellingARMv8}, the behaviour that we allow was never observed, but it is also allowed by the model
in \cite{HerdingCats}.  
%As such we deem this discrepancy to be a minor issue in Flowing/POP (preservation of transitive dependencies) rather than a fault in our model.

\OMIT{
%%%%%%%%%%%%%%%%%%%%%%%%%%%%%%5
We have renamed some variables and simplified some of the syntax for clarity, in particular introducing a higher-level $\If$ statement to remove a jump
command and implicit register (referenced by the compare (\code{CMP}) and branch-not-equal (\code{BNE}) instructions).
The key point is that the expression of the $\If$ statement references register \code{r3} which holds the loaded value of $z$.  Hence the branch cannot be
evaluated (non-speculatively) until the value for $z$ is known.
\begin{equation}
\begin{split}
\label{eq:ppo015}
	~&
 \storex{1}
 \cbef
 \fence
 \cbef
 \storey{1}
 % \\
 	% ~&
 \quad
 \pl
 \\
 	~&
 \load{r0}{y}  
 \cbef
  \mathttbf{eor}~\T{r1, r0, r0}
 \cbef
  \mathttbf{add}~\T{r1, r1, 1}
 \cbef
  \store{z}{r1}
 \cbef
  \store{z}{2}
 \cbef
  \load{r3}{z}
 \cbef
  \\ 
  	~&
  \qquad \qquad
  (\If \T{r3 = r3} \Then \Skip \Else \Skip)
 \cbef
  \cfence
 \cbef
  \load{r4}{x}
\end{split}
\end{equation}

The tested condition is 
$z=2 \land r0=1 \land r4=0$, which asks whether it is possible to load $x$ (the last statement of process 2) before loading $y$ (the first statement of
process 2).

At a first glance the control fence prevents the load of $x$ happening before the branch.  However, litmus tests such as
\T{MP+dmb.sy+fri-rfi-ctrisb}, \citep[Sect 3,\emph{Out of order execution}]{ModellingARMv8}, indicate that under some circumstances the branch condition can be
evaluated early.  In our model, the instruction $\load{r3}{z}$ can be reordered before $\store{z}{2}$, which forwards the value 2 to \code{r3}.  Now
$\reg{r3}{2}$ can be executed (reordered before all earlier instructions), and subsequently so can the branch instruction $\guard{r3 = r3}$. 
This allows the control fence to be executed since there is no preceding branch point, 
and finally $\load{r4}{x}$, since it is not dependent on any earlier register or shared variable and is no longer blocked by the control fence.  
As such the value loaded into $r4$ may be the initial value for $x$.  Process 1 then executes to completion, after which process 2 continues by loading the value 1 for
$y$.

In the Flowing/POP model of \citep{ModellingARMv8}, 
this behaviour is forbidden because there is a data dependency from the load of $y$ into \code{r0} to \code{r3}, via $z$.  This
appears to be because of the consecutive stores to $z$, one of which depends on the value in \code{r1} which transitively depends on \code{r0}.  

In the testing of real processors reported in \citep{ModellingARMv8}, the behaviour that we allow was never observed.  This may because that behaviour
is intended to be allowed by the specification of the architecture, but is not realised on any processor.  However, it may also be intended to be
disallowed, in which case our model is wrong.

If the behaviour of litmus test \T{PPO017} \refeqn{ppo017} is intended to be forbidden, that requires forwarding values only when earlier stored values have been fully
determined (i.e., when the value for \T{r1} is known).  That cannot be determined with the simple pair-wise comparison that our framework uses.
It would also appear to limit the amount reordering possible, for no change in sequential semantics (it is still valid to forward the known
value 2 for $z$ to later loads of $z$).  As such we deem this discrepancy to be a minor issue in Flowing/POP (preservation of transitive
dependencies) rather than a fault in our model.
}

\section{POWER}
\label{power}

\newcommand{\lwflush}{lwflush}
\newcommand{\seesnm}[1]{\sees{\pidn}{\pidm}{#1}}
\newcommand{\seesnmwt}{\seesnm{\wseq_2}}
\newcommand{\seesnmw}{\seesnm{\wseq}}

POWER processors  allow similar reorderings to ARM, as well as those aspects discussed in
\refsect{speculative-execution} to \refsect{elimEarlierWrites}.  Additionally, POWER includes so-called
``lightweight fences'', $\lfence$, which have both a local and global effect, and the similar $\eieio$ barriers.
The reordering and forwarding definitions for POWER are otherwise the same as those in \reffig{higher-level-defs}; we discuss how that relates to hardware tests in \refsect{sec:power-validation}.

\subsection{Lightweight fences}

A lightweight fence, denoted $\lfence$, maintains order between loads, loads then stores, and stores, but not stores and subsequent loads
(i.e., \T{load;load}, \T{load;store}, \T{store;store}, but not \T{store;load}).  
As shown in the reordering rule \refeqn{lfence-def} of \reffig{lwfence-semantics} we use two types of instruction to model the lightweight fence instruction.%
\footnote{
If lightweight fences did not maintain
\T{load;load} order it would be straightforward to define their effect in terms of one instruction.  
It is possible to do it with one fence that marks any later load which `jumps' it so that load can be
reordered with earlier stores but \emph{not} earlier loads, however this seems less elegant and less 
amenable to algebraic analysis.
}
These are ``gates'', with the $\storegate$ instruction allowing stores to ``move backwards'' and the $\loadgate$ instruction allowing loads to ``move forward'', as
given by the reordering rules \refeqns{storegate-ro}{loadgate-ro}.

For instance, assume the following sequence of instructions, where $l_i$ are loads and $s_i$ are stores.
\[
	l_1 \cbef s_1 \cbef \storegate \cbef \loadgate \cbef l_2 \cbef s_2
\]
Assuming all loads and stores are to different variables and hence there are no pairwise constraints on reordering,
application of the \reflaws{keep-order}{swap-order} gives
\[
	l_1 \bef \storegate \bef l_2 \bef s_1 \bef \loadgate \bef s_2
\]
Note that the order between loads, between stores, and between loads then stores has been maintained, but load $l_2$ may be reordered before the store $s_1$.

In addition to a local constraint on possible reorderings,
a lightweight fence also has a global effect on the storage system, which we encode in the semantics of the
$\loadgate$ instruction (\refrule{storage-loadgate}).  Informally, a lightweight fence requires other processes to see changes to (different) shared variables in the same order as the process
that wrote them.  This is subtly different to a full fence, as we describe below.

\paragraph{Effect on the storage subsystem}

As given in \refrule{storage-loadgate}
a $\loadgate$ instruction by process $\pidn$ has an effect on the storage $\wseq$ similar to a full fence, but in this case all writes that $\pidn$ has seen (which
includes all writes by $\pidn$) are also tagged as lightweight-fenced by $\pidn$ by adding the $\lwfencedn$ tag to the $seen$ part of a write.
This is given by the recursive definition of $\lwflush$ in \refeqn{def-lwflush}.

In the presence of lightweight fences the rule for a load (\refrule{storage-load}) changes to \refrule{storage-load-lwf}.
The change is that if process $\pidn$ sees a write by process $\pidm$ then it must also see any earlier writes by $\pidm$ that $\pidm$ has lightweight-fenced.  
This is given by the function $\seesnm{\wseq}$ which simply marks any writes that are lightweight-fenced by $\pidm$ to be seen and lightweight-fenced by $\pidn$.
This transitive effect gives cumulativity of lightweight fences \citep{HerdingCats}. 

The inclusion of a lightweight fence also has a subtle effect on the write order in the storage subsystem: the definition of write reordering as used in
\refrule{storage-store} is updated to \refeqn{canReorder-lwf}, where
$\canReorder{w_1}{w_2}$ is defined so that write $w_2$ by $\pidn$ to $x$ may come before write $w_1$ (to different variable $y$) provided that $w_1$ has not previously
been lightweight-fenced by $\pidn$.

\begin{figure}

\begin{eqnarray}
	\aca &\ttdef& \ldots \csep \storegate \csep \loadgate \csep \eieio
	\label{loadgate-syntax}
	\label{eq:storegate-syntax}
	\\
	\lfence \cbef c
	&\sdef& 
	\storegate \cbef \loadgate \cbef c
	\label{eq:lfence-def}
	\\
		\aca &\ro& \storegate 
		\qquad 
		\ltif{$\aca$ is a store}
\label{eq:storegate-ro}
		\\
		\loadgate &\ro& \aca
		\qquad 
		\ltif{$\aca$ is a load}
\label{eq:loadgate-ro}
	\\
	\eieio 
	&\nro&
	\aca
		\ltif{$\aca$ is a store}
\label{eq:e<a}
	\\
	\aca
	&\nro&
	\eieio 
		\ltif{$\aca$ is a store}
\label{eq:a<e}
\end{eqnarray}

\ruledefNamed{0.98\textwidth}{Storage - lightweight fence}
{storage-loadgate}{
	\Rule{
		p \ttra{\pidn:\loadgatelbl} p'
	}{
		\storage{\wseq}{p}
		\ttra{\pidn:\loadgatelbl}
		\storage{\lwflush_\pidn(\wseq)}{p'}
	}
}

\ruledefNamed{0.99\textwidth}{Storage - load in presence of lightweight fences}
{storage-load-lwf}{
	\Rule{
		p \ttra{\pidn:\Readxv} p'
		\also
		\all w \in \ran(\wseq_2) @ \canSeePastnxw
		%\qquad
		%\wseq_2' = \sees{\pidn}{\pidm}{\wseq_2}
		%%\all w \in \ran(\wseq_1) @ x = w.var \imp \pidn \notin w.seen 
	}{
		\begin{array}{l}
		\storage{\wseq_1 \scat \srqmxv{\pidns} \scat \wseq_2}{p}
		\ttra{\pidn:\Readxv}
		\also
		%\qquad \qquad \qquad \qquad \qquad \qquad
		\storage{\seesnm{\wseq_1} \scat \srqmxv{\pidns \union \{\pidn\}} \scat \wseq_2}{p'}
		%\storage{\wseq_1 \scat w \scat \wseq_2}{p}
		%\ttra{\pidn:\Readxv}
		%\storage{\wseq_1 \scat w' \scat \wseq_2}{p'}
		\end{array}
	}
}

\OMIT{
\ruledefNamed{0.99\textwidth}{Storage - store}
{storage-store}{
	\Rule{
		p \ttra{\pidn: x \asgnlbl v} p'
		\\
		w' = \srqnxvn
		\qquad
		\all w \in \ran(\wseq_1) @ \canReorder{w}{w'}
		%\all w \in \ran(\wseq_1) @ \canReorder{w}{\srqnxvn}
		%% \all w \in \ran(\wseq_1) @ 
			%% \pidn \neq w.thread 
			%% \land 
			%% (x = w.var \imp \pidn \notin w.seen)
	}{
		\storage{\wseq_1 \scat \wseq_2}{p}
		\ttra{\pidn: x \asgnlbl v}
		\storage{\wseq_1 \scat w' \scat \wseq_2}{p'}
		%\storage{\wseq_1 \scat \srqnxvn \scat \wseq_2}{p'}
	}
}
}

where
\begin{eqnarray}
	\Also
	\lwflush_\pidn(\eseq) = \eseq
	&&
	\lwflush_\pidn(\wseq \scat w) = 
	\lwflush_\pidn(\wseq) \scat \lwflush_\pidn(w)
	\notag
	\\
	%\lwflush_\pidn(\srqmxvns \scat \wseq) &=& 
	\lwflush_\pidn(\srqmxvns) &=& 
	\left\{
		\begin{array}{ll}
		\srq{\pidm}{x}{v}{\pidns \union \{\lwfenced{\pidn}\}} 
			%\scat \lwflush_\pidn(\wseq) 
		%\\&&
		%\notag
		&
		if~ \pidn \in \pidns
	\\
	%\lwflush_\pidn(\srqmxvns \scat \wseq) &=& 
		\srqmxvns 
			%\scat \lwflush_\pidn(\wseq) 
		%\\&&
		&
		\mbox{otherwise}
		%if~ \pidn \notin \pidns
	\end{array}
	\right.
	\label{eq:def-lwflush}
	%\canSeePastnxw &\sdef& x = w.var \imp \pidn \notin w.seen 
	%\\
	\\
	\Also
	\seesnm{\eseq}
	= \eseq
	&&
	\seesnm{\wseq \scat w} = \seesnm{\wseq} \scat \seesnm{w}
	\notag
	\\
	%\seesnm{\srq{-}{x}{v}{\pidns} \scat \wseq}
	\seesnm{\srq{-}{x}{v}{\pidns}}
	&=& 
	\left\{
	\begin{array}{ll}
		%\Update{w}{seen}{\pidns \union \pidn}
		\srq{-}{x}{v}{\pidns \union \{\lwfencedn, \pidn\}}
			%\scat \seesnmw
		&
		\mbox{if $\lwfenced{\pidm} \in \pidns$}
		%\mbox{if $w.seen = \pidns$ and $\lwfenced{\pidm} \in \pidns$}
		\\
		\srq{-}{x}{v}{\pidns}
			%\scat \seesnmw
		&
		\mbox{otherwise}
	\end{array}
	\right.
	\\
	\canReorder{w}{\srqnxvn} 
	&\sdef&
			\pidn \neq w.thread 
			\land 
			\notag \\ && \qquad
			(x = w.var \imp \pidn \notin w.seen)
			\land 
			\label{eq:canReorder-lwf}
			\\ && \qquad \notag
			(x \neq w.var \imp \lwfenced{\pidn} \notin w.seen)
\end{eqnarray}

\caption{Additions to storage subsystem to handle POWER's lightweight fences}
\label{fig:lwfence-semantics}
\end{figure}

%\subsection{\T{eieio} barriers}
%\label{eieio}

In addition to lightweight fences POWER also includes an \T{eieio} barrier.
Based on the discussion in \cite{HerdingCats} we treated this as a barrier on stores only \refeqns{e<a}{a<e}.
In addition an $\eieio$ barrier acts as a lightweight flush on the storage system as in \refrule{storage-loadgate}.
%This interpretation agrees with the model of \cite{HerdingCats} in all 583 tests we ran.

\subsection{Validation}
\label{sec:power-validation}

There are two litmus test resources we used.
Firstly,
we validated the semantics against a set of 758 litmus tests taken from the supplementary material of \citep{UnderstandingPOWER}%
\footnote{
\url{https://www.cl.cam.ac.uk/~pes20/ppc-supplemental/index.html}
}
(the reason for more tests is due to the extra $\lfence$ cases).  As with ARM, some tests were excluded
due to parsing issues.

As with the results from \cite{ModellingARMv8}, our model disagrees on the same three tests discussed in \refsect{ppo017}, 
and otherwise agrees with the expected results on hardware, and agrees with
the results of their model where available (52 litmus tests did not complete
in a reasonable amount of time using the tool in \citep{UnderstandingPOWER}).\footnote{
Litmus test \T{propagate-sync-coherence} was apparently not run on hardware, but our model's result agreed with that of \citep{UnderstandingPOWER}.
}

%One of the litmus tests (\T{IRIW+addrs-twice.litmus}) did not complete within a reasonable amount of time in our tool; that was also one of 52 that did not
%complete within a reasonable amount of time using the tool in \citep{UnderstandingPOWER}.

The second set of tests comes from \cite{HerdingCats}.  As we maintain the same ordering relationship for ARM and POWER, yet the POWER model in \cite{HerdingCats} was
stronger, unsurprisingly our model disagrees with more cases for POWER.  That is, of the 7820 successful runs (approx. 350 tests were excluded due to parsing or
timeout problems) 550, or 7\%, disagreed with the model in \cite{HerdingCats} or the hardware results.%
\footnote{
We note that the POWER model of \HC allows behaviours not observed in 1002 out of 8141 tests;
there are no cases where the model of \HC forbids behaviour that was observed.
}
Based on the discussion in \cite{HerdingCats}[Sect. 8.1] we conjecture that this is because forwarding, especially with respect to
branches, is handled differently.  However, given the conformance with the more recent 758 tests reported above, and for reasons discussed in \refsect{validation-overview}, we have not pursued a specific change to our model to
account for the discrepancies.

Instead we note that if we exclude litmus tests with multiple stores to the same variables in one process, multiple loads of the same variable in one process (i.e.,
\T{po-loc} issues), 
there are only 18 tests out of 3142 where our model disagrees with both the model in \cite{HerdingCats} and the POWER hardware they tested.
%\rcnote{Or we could say 18 out of 3142 if we include address shifting}

%% \begin{verbatim}
	%% ./parse-results.py -hpower -norepeatFailed | grep -v 'StoreStore' | grep -v 'missing L' | grep -v Quantifi | grep -v HH | grep -v PROBLEM  | wc -l
%% \end{verbatim}

\section{Verification of higher-level algorithms}
\label{higher-level-code}

We now show how our semantics and its Maude encoding can be used to investigate running programs expressed in our wide-spectrum language on architectures with weak memory models.
Throughout we assume that simple assignments are atomic, noting where we use more
complex constructs such as compare-and-swap and allocating a new node on the heap.
The code listings are close to the Maude code used for verification, but are refactored from the sources in the literature to eliminate returns
from inside a loop or branch, etc., so that the code is expressed in a straightforward structure that preserves the intended paths.

\subsection{Locking}

We analyse a simple lock/unlock algorithm for correctness under ARM and POWER (the code is
taken from a version for Java in \citep[Sect. 7.3]{HerlihyShavit2011}).

\OMIT{
\begin{lstlisting}[multicols=2]
	initially locked $\asgn$ false

	unlock $\sdef$ 
		locked $\asgn$ false

lock $\sdef$ 
	locals success $\asgn$ false
	while $\neg$success do 
		success $\asgn$ $\neg$locked.getAndSet(true)
\end{lstlisting}
}
\hspace{-15mm}
\parbox{0.48\textwidth}{
\[
\noindent
Initial~state: \{locked \mapsto false\}
\\~\\
unlock \sdef \\
  \quad locked \asgn false \\
\]
}
\hspace{-10mm}
\parbox{0.60\textwidth}{
\[
\noindent
~\\
~\\
lock \sdef \\
  \quad \LOCALSWord~ success \mapsto false @ \\
  \quad \WHILEWord \neg success \DOWord \\
  \quad \quad success \asgn \neg locked.getAndSet(true) 
\]
}

We use $\LOCALSWord$ to declare and initialise the local variable names used (registers in assembler code).
An important part of the algorithm is that the \code{unlock} operation does not need to include a fence.
The key line of code is 
\code{
	success \asgn \neg locked.getAndSet(true) 
}
which in our framework we treat as an atomic block with an implicit fence.
\[
	success \asgn \neg locked.getAndSet(true) \sdef
	\\ \qquad \qquad
 		\atomic{ \guard{\neg locked} \atomicsep locked \asgn true \atomicsep \fence } \cbef success \asgn true 
		%\\ \qquad \qquad
		\quad
		\choice
		\\ \qquad \qquad
   		\guard{locked} \cbef success \asgn false 
\]
The primitive \code{locked.getAndSet(b)} atomically changes the boolean variable \code{locked} to $b$ and returns the previous value of \code{locked}.  
Thus if \code{locked} is
already true \code{locked.getAndSet(true)} has no effect and returns true; if \code{locked} is false then it becomes true and false is returned.
The atomic action may be implemented using load-linked/store-conditional or compare-and-swap primitives, but we abstract from that and use an atomic block to define
\code{getAndSet}.
Note that the guard condition references a global variable.  This is not possible in ARM or POWER directly, requiring a load of \code{locked} into a local register first,
however we can straightforwardly accommodate this in our general semantics. 

%Theoretically a loop can be speculatively executed infinitely far ahead, however in this case the body of the loop (containing a store and a load) modifies the
%variable in the loop condition and hence speculation is not possible.

%A similar algorithm is tested in \cite{ARMv8.4}, taken from the linux kernel.  We cannot directly test that algorithm as it involves low-level instructions
%that we have not modelled (though presumably could).  

\subsubsection{Model checking}

To test this lock implementation provides mutual exclusion in ARM or POWER, we call \code{lock ; unlock} with an intervening
abstract critical section, in which a flag per process is set to true/false on entry/exit, with one process setting a variable \T{conflict} to true if both are in
their critical section at the same time.
For model checking purposes we unfold the loop a finite number of times, giving a structure of nested branches.  Many of the generated
paths end with a guard testing \code{success}, which may be false; in which case that path is removed from the analysis.

The prototype tool was able to confirm mutual exclusion was satisfied for two concurrent processes with approximately 800 million rewrites in 14 minutes of processor time.  
%A similar test using three processes did not complete within a reasonable timeframe.

\subsection{Treiber stack}

\newcommand{\stack}{s}

We model the Treiber stack, a well known lock-free data structure implementing an abstract stack \citep{Treiber86}, in which \code{push} and \code{pop} operations can be called concurrently.
The operations are structured as potentially infinite loops which retry if interference occurs on the \code{Head} of the stack, otherwise succeeding and modifying \code{Head}
atomically with a compare-and-swap instruction.

\OMIT{
\begin{lstlisting}[multicols=2]
push(V) $\sdef$
	locals head $\asgn$ nullPtr n $\asgn$ nullPtr
	n $\asgn$ new Node(V) 
	repeat
		head $\asgn$ Head 
		*n.next $\asgn$ head 
	until CAS(Head, head, n) 




pop $\sdef$ 
	locals head $\asgn$ nullPtr, n $\asgn$ nullPtr, return $\asgn$ retry
	while return = retry do 
		head $\asgn$ Head 
		if head = nullPtr then
			return $\asgn$ emptyStack 
		else 
			n $\asgn$ heap(head) 
			if CAS(Head, head, n.next) then
				return $\asgn$ (n.value) 
\end{lstlisting}
}

\newcommand{\nullPtr}{\mathop{\mathsf{null}}}
\newcommand{\retry}{\mathop{\mathsf{retry}}}
\newcommand{\emptyStack}{\mathop{\mathsf{empty}}}
\newcommand{\NEWWord}{\mathop{\mathbf{new}}}
\newcommand{\REPEATWord}{\mathop{\mathbf{repeat}}}
\newcommand{\UNTILWord}{\mathop{\mathbf{until}}}
\newcommand{\nodeCons}{\mathop{\mathsf{node}}}

\hskip-15mm
\parbox[t]{0.46\textwidth}{
\[
\noindent
Initial~state: \{head \mapsto \nullPtr\} 
\]
}

\hskip-15mm
\parbox[t]{0.46\textwidth}{
\[
push(v) \sdef \\
  \quad \LOCALSWord~ 
	head \mapsto \nullPtr, n \mapsto \nullPtr
  @ \\
	\quad 
	n \asgn \NEWWord Node(v) \\ 
	\quad
	\REPEATWord \\
		\quad\quad
		head \asgn Head \\
		\quad\quad
		\*n.next \asgn head  \\
	\quad
	\UNTILWord CAS(Head, head, n) 
\]
}
\hskip-5mm
\parbox[t]{0.63\textwidth}{
\[
\noindent
pop \sdef \\
  \quad \LOCALSWord~ 
	head \mapsto \nullPtr, n \mapsto \nullPtr, \\
		\quad\quad\quad\quad
		return \mapsto \retry
  @ \\
  	\quad
	\WHILEWord return = \retry \DOWord  \\
		\quad\quad
		head \asgn Head  \\
		\quad\quad
		\IFWord head = \nullPtr \THENWord \\
			\quad\quad\quad
			return \asgn \emptyStack  \\
		\quad\quad
		\ELSEWord \\
			\quad\quad\quad
			n \asgn heap(head)  \\
			\quad\quad\quad
			\IFWord CAS(Head, head, n.next) \THENWord \\
				\quad\quad\quad\quad
				return \asgn n.value 
\]
}

The algorithm accesses the heap,
which we model as individual shared variables (addresses) $heap(0)$, $heap(1)$, etc., which can be both assigned to and appear in expressions.  
An implicit shared variable \code{maxh}, initially 0, keeps track of the next free address in the heap. Addresses are allocated in order from 0.
A node value is the term \code{\nodeCons(v, p)} where \code{p} is a pointer (either a natural number index into the heap or \code{\nullPtr}).%
\footnote{
If pointers are freed and reused the algorithms suffer from the common ``ABA problem'', where a new node may reuse a heap location and cause a \code{CAS} operation to
succeed where it should fail.  The chances of this occurring in a practical setting can be made acceptably small by introducing modification counts to the pointers
\citep{Moir97}.  For our model checking
purposes we assume the system provides some garbage collection mechanism that avoids this problem and we do not explicitly model freeing nodes or avoiding the ABA problem.
}

We assume the underlying system provides the following implementation of
$n \asgn \NEWWord Node(v)$.
\begin{equation*}
\atomic{ heap(maxh) \asgn \nodeCons(v, \nullPtr) \atomicsep n \asgn maxh \atomicsep maxh \asgn (maxh + 1) \atomicsep \fence }
\end{equation*}
That is, when a new node is allocated,
the next available address is initialised to a new node value, \code{maxh} is incremented, the return value \code{n} is set to point to the new allocation, and a
final $\fence$ ensures that all other threads see the effect on \code{maxh} atomically.\footnote{
Note that the $\fence$ does not affect variables other than \code{maxh} as there are no preceding instructions in \code{push} and all operations (potentially preceding the \code{push})
end with a $\CASWord$ with an implicit fence \refeqn{tso-cas}.
}

Updating the \code{next} or \code{val} part of a node also uses a shorthand, i.e.,
\code{n.next \asgn head}
is expanded to 
%\begin{equation}
%\mbox{
\code{
heap(n) \asgn \nodeCons(heap(n).value, head) 
},
%}
%\end{equation}
which overwrites the node's next pointer, but keeps the original value.

\subsubsection{Verification}
\label{treiber-verification}

To verify this algorithm we compare the final values of the stack and the return values of the \code{pop} operations against those of an abstract specification of a
stack (we do not prove termination).
We use braces \code{\lseq..\rseq} to enclose sequences and $\scat$ for sequence concatenation.
\begin{eqnarray}
	push(v) &\sdef& \stack \asgn \seqT{v} \scat \stack
	\\
	pop &\sdef& 
		\LOCALSWord return \mapsto \emptyStack @ 
		\\ &&
		    \atomic{ \guard{\stack = \eseq} \atomicsep return \asgn \emptyStack }   \choice
			\\ &&
		    \atomic{ \guard{\stack \neq \eseq} \atomicsep return \asgn head(\stack) \atomicsep \stack \asgn tail(\stack) }
\end{eqnarray}
The operations are modelled using atomic steps, and we model an explicit return value in the abstract
specification.

%Because we check the values returned by \code{pop} operations our verification is similar to that of linearisability \citep{linearisability}.  Informally, for an
%operation to be linearisable it must appear to take effect atomically.  Linearisability is formally defined with respect to sequences of operation
%invocation/response events, where the concrete sequences must be transformable (according to certain rules) into an equivalent sequence generated by an atomic
%specification.  In our setting, the return values can take the place of the \code{pop} response actions.  However, 
Although we check the values returned by \code{pop}, we do not attempt to relate intermediate values of
the stack to their abstract counterpart, only the final value.  Agreement on final values of the stack does give some confidence that incorrect values are not
discovered during the execution; however we emphasise that this approach is an approximation of correctness only.  Nevertheless, 
the technique can potentially
identify problems in running algorithms on weak memory models: the same technique exposed a flaw in a published algorithm, as discussed in \refsect{chase-lev}.

We validate the code against the abstract specification by running a process $p$ which is a combination of push and pop operations, for instance \code{push(v)
\pl pop}.  We 
%collected the final values of the stack and the return values from the \code{pop} operations.  We then ran \code{push(v) \pl pop} and 
then check
that the heap and \code{Head} pointers abstractly give a valid stack, and the return values correspond with the expected abstract return values.

%We keep the loops small by limiting the number of iterations, however this is not that relevant as there are only as many times through the loops as there are processes, due
%to interference (assuming they are correct wrt. lock-freedom).

We ran 4 combinations of parallel processes (in addition to testing \code{push} and \code{pop} running in isolation), and all four gave the expected abstract behaviour.  
The most time intensive test involved three concurrent processes formed from a \code{push} and two \code{pop}s.  This required 103 billion rewrites and took 23 hours to return.  
These results give confidence that the Treiber stack algorithm will work on ARM and POWER-style weak memory models assuming \code{CAS} is implemented as in
\refeqn{tso-cas}.

\subsection{Chase-Lev deque}
\label{chase-lev}

\newcommand{\none}{\mathop{\mathsf{none}}}
\newcommand{\emptyDeque}{\mathop{\mathsf{empty}}}
\newcommand{\fail}{\mathop{\mathsf{fail}}}

L\^{e} et. al \citep{LeWorkStealingPPoPP13} present a version of the  
Chase-Lev deque (double-ended queue) \citep{ChaseLev05} adapted for weak memory models.
The deque is implemented as an array, where elements may be \emph{put} on or \emph{taken} from the tail,
and additionally, processes may \emph{steal} an element from the head of the deque.
The \code{put} and \code{take} operations
may be executed by a single thread only, hence there is no interference between these two operations, although the thread-local reorderings could cause consecutive
invocations to overlap.  The \code{steal} operation can be executed by other processes.

The code we test is given in \reffig{chaselev-code}.
The original code includes handling array resizing, but here we focus on the insert/delete logic.
As above, we have refactored the algorithm to eliminate returns from within a branch, and use \code{CAS} terminology for the atomic updates.

\OMIT{
\begin{figure}
\begin{lstlisting}
	initially head $\asgn$ 0, tail $\asgn$ 0, tasks $\asgn$ [_,_,...]
	put(v) $\sdef$
		locals t $\asgn$ 0
		t $\asgn$ tail
		tasks[t mod L] $\asgn$ v
		fence 
		tail $\asgn$ t + 1
\end{lstlisting}

\begin{lstlisting}[multicols=2]
	take $\sdef$
		locals h $\asgn$ 0, t $\asgn$ 0, return $\asgn$ 0
		t $\asgn$ tail - 1
		tail $\asgn$ t
		fence 
		h $\asgn$ head
		if h <= t then
			return $\asgn$ tasks[t mod L]
			if h = t then
				if !CAS(head, h, h + 1) then
					return $\asgn$ empty 
				tail $\asgn$ (t + 1) 
		else
			return $\asgn$ empty 
			tail $\asgn$ (t + 1)

	steal $\sdef$
		locals h $\asgn$ 0, t $\asgn$ 0, task $\asgn$ 0, return $\asgn$ _
		h $\asgn$ head
		fence 
		t $\asgn$ tail
		cfence // unnecessary
		if h < t then
			task $\asgn$ tasks[h mod L]
			cfence // incorrectly placed.
			if CAS(head, h, h+1) then
				return $\asgn$ task
			else
				return $\asgn$ fail
		else
			return $\asgn$ empty
\end{lstlisting}
\caption{A version of L\^{e} et. al's work-stealing deque algorithm for ARM}
\label{fig:chaselev-code}
\end{figure}
}

\begin{figure}[tp]

\hspace{-5mm}
\parbox{0.87\textwidth}{
\[
\noindent
Initial~state: \{head \mapsto 0, tail \mapsto 0, tasks \mapsto \seqT{\_, \_, \ldots}\}
\\~\\
put(v) \sdef \\
  \quad \LOCALSWord~ t \mapsto \_ @ \\
  \quad t \asgn tail \cbef \\
  \quad tasks[t \mod L] \asgn v \cbef \\
  \quad \fence \cbef \\
  \quad tail \asgn t + 1
\]
}
\vskip -5mm
\hspace{-5mm}
\parbox{0.495\textwidth}{
\[
 take \sdef\\
	\quad \LOCALSWord~ h \mapsto \_ , t \mapsto \_ , return \mapsto \_ @ \\
	\quad t \asgn tail - 1 \scomp \\
	\quad tail \asgn t \scomp \\
	\quad \fence  \scomp \\
	\quad h \asgn head \scomp \\
	\quad \IFWord h \leq t \THENWord \\
	  \quad \quad return \asgn tasks[t \mod L] \scomp \\
	  \quad \quad \IFWord h = t \THENWord \\
	  	%\quad \quad \quad \IFWord \neg \CAS{head}{ h}{ h + 1} \THENWord \\
	  	\quad \quad \quad \IFWord \neg \CAS{head}{ h}{ h + 1} \\
	  	\quad \quad \quad \THENWord \\
	  		\quad \quad \quad \quad return \asgn \emptyDeque  \\
	  	\quad \quad \quad tail \asgn t + 1  \\
	\quad \ELSEWord \\
	  \quad \quad return \asgn \emptyDeque  \scomp \\
	  \quad \quad tail \asgn t + 1 \\
\]
}
\hspace{-5mm}
\parbox{0.53\textwidth}{
\[
 steal \sdef \\
	\quad \LOCALSWord~ h \mapsto \_ , t \mapsto \_ , return \mapsto \_  @ \\
	\quad h \asgn head \scomp \\
	\quad \fence  \scomp \\
	\quad t \asgn tail \scomp \\
	\quad \cfence \scomp \color{dkgreen}{//~unnecessary}  \\
	\quad \IFWord h < t \THENWord \\
	  \quad \quad return \asgn tasks[h \mod L] \scomp \\
	  \quad \quad \cfence \scomp \color{dkgreen}{//~incorrectly~placed} \\
	  \quad \quad \IFWord \neg \CAS{head}{ h}{ h+1} \THENWord \\
	  	  %\quad \quad \quad return \asgn task \\
	  %\quad \quad \ELSEWord \\
	  	  \quad \quad \quad return \asgn \fail \\
	\quad \ELSEWord \\
	  \quad \quad return \asgn \emptyDeque
	\\~\\
	~\\
\]
}

\caption{A version of L\^{e} et. al's work-stealing deque algorithm for ARM \cite{LeWorkStealingPPoPP13}}
\label{fig:chaselev-code}
\end{figure}

The \code{put} operation straightforwardly adds an element to the end of the deque, incrementing the \code{tail} index.  It includes a full fence so that the tail
pointer is not incremented before the element is placed in the array.  The \code{take} operation uses a \code{CAS} operation to atomically increment the head index.
Interference can occur if there is a concurrent \code{steal} operation in progress, which also uses \code{CAS} to increment \code{head} to remove an element from the
head of the deque.  The \code{take} and \code{steal} operation return empty if they
observe an empty deque.  In addition the \code{steal} operation may return the special value $\fail$ if interference on \code{head} occurs.  Complexity arises if the
deque has one element and there are concurrent processes trying to both \code{take} and \code{steal} that element at the same time.  

Operations \code{take} and \code{steal} use a $\fence$ operation to ensure they have consistent readings for the head and tail indexes, and later use \code{CAS} to
atomically update the head pointer (only if necessary, in the case of \code{take}).  Additionally, the \code{steal} operation contains two $\cfence$ barriers
(\T{ctrl\_isync} in ARM).  Our analysis suggests that
the first control fence is redundant, and the second
is incorrectly placed.  Eliminating the first $\cfence$ and swapping the order of the second control fence with the preceding load into \code{task} 
gives the expected behaviour.
We describe this in more detail below.

\subsubsection{Verification}
\label{chaselev-verification}

\def\absput{put}
\def\abstake{take}
\def\abssteal{steal}

As with the Treiber stack we use an abstract model of the deque and its operations to generate the allowed final values of the deque and return values.
The function \code{last(q)} returns the last element in \code{q} and \code{front(q)} returns \code{q} excluding its last element.
\OMIT{
\begin{lstlisting}
	put(v) $\sdef$
		q $\asgn$ q $\scat$ [v] 
	abs-take $\sdef$
		locals return $\asgn$ none
		    < guard(q = {}) / return $\asgn$ empty >  []
		    < guard(q $\neq$ {}) / return $\asgn$ last(q) / q $\asgn$ front(q) >
	abs-steal $\sdef$
		locals return $\asgn$ none
		    < guard(q = {}) / return $\asgn$ empty >  []
		    < guard(q $\neq$ {}) / return $\asgn$ head(q) / q $\asgn$ tail(q) >
\end{lstlisting}
}
\begin{eqnarray*}
	\absput(v) &\sdef&
		q \asgn q \cat \seqT{v}
	\\
	\abstake &\sdef&
		\LOCALSWord\ return \asgn \none @ 
			\\ && \quad
		    \atomic{ \guard{q = \eseq} \atomicsep return \asgn \emptyDeque }  \choice
			\\ && \quad
		    \atomic{ \guard{q \neq \eseq} \atomicsep return \asgn last(q) \atomicsep q \asgn front(q) }
	\\
	\abssteal &\sdef&
		\LOCALSWord\ return \asgn \none
			\\ && \quad
		    \atomic{ \guard{q = \eseq} \atomicsep return \asgn \emptyDeque } \choice
			\\ && \quad
		    \atomic{ \guard{q \neq \eseq} \atomicsep return \asgn head(q) \atomicsep q \asgn tail(q) }
\end{eqnarray*}

The abstract specification for \code{steal} is not precise as it does not attempt to detect interference and return $\fail$.  As such we exclude these behaviours of
the concrete code from the analysis.

That the first control fence is redundant is shown by the following derivation.
\begin{derivation}
	\step{
		h \asgn Head \cbef \fence \cbef t \asgn Tail \cbef \cfence \cbef \If \ldots
	}
	\trans{\refsto}{\reflaw{keep-order}}
	\step{
		h \asgn Head \cbef \fence \cbef  t \asgn Tail \bef \cfence \cbef\If \ldots
	}
	\trans{\refsto}{\reflaw{swap-order}}
	\step{
		h \asgn Head \cbef \fence \cbef \cfence \bef t \asgn Tail \cbef \If \ldots
	}
\end{derivation}
The control fence can be reordered before the previous load; it is now immediately after a fence, and has no further effect on reorderings of later
instructions, and hence is redundant.

Our model checking using the Maude encoding exposes a bug in the code which may occur when a \code{put} and \code{steal} operation
execute in parallel on an empty deque.  The load $task \asgn tasks[h \mod L]$ can be speculatively executed before the branch is evaluated, and hence also
before the load of \code{tail}.  Thus the steal process may load \code{head}, load an irrelevant value for \code{task}, at which point a \code{put} operation
may complete, storing a value and incrementing \code{tail}.  The \code{steal} operation resumes, loading the new value for \code{tail} and observing a
non-empty deque, succeeding with its \code{CAS} and returning the irrelevant value in \code{task}, which was loaded before the \code{put} operation had
begun.

More concretely, the following reordering is possible (similar to the derivation in \refsect{speculative-execution}).
We have removed the first (redundant) control fence, and we leave much of the structure summarised as $\ldots$ as it is only the first instructions that are
relevant.
\begin{derivation}
	\step{
		h \asgn Head \cbef \fence \cbef t \asgn Tail \cbef 
	}
	\step{
		\qquad\qquad
		\If h < t \Then 
			return \asgn tasks[h \mod L] \cbef \cfence \ldots
	}

	\trans{\refsto}{Defn. of $\If$; resolve to first branch}
	\step{
		h \asgn Head \cbef \fence \cbef t \asgn Tail \cbef
	}
	\step{
		\qquad\qquad
		\guard{h < t} \cbef
			return \asgn tasks[h \mod L] \cbef \cfence \ldots
	}

	\trans{\refsto}{\reflaw{keep-order}}
	\step{
		h \asgn Head \cbef \fence \cbef t \asgn Tail \cbef
	}
	\step{
		\qquad\qquad
		\guard{h < t} \cbef
			return \asgn tasks[h \mod L] \bef \cfence \ldots
	}

	\trans{\refsto}{\reflaw{swap-order}}
	\step{
		h \asgn Head \cbef \fence \cbef t \asgn Tail \cbef
	}
	\step{
		\qquad\qquad
		return \asgn tasks[h \mod L] 
		\bef
		\guard{h < t} \cbef
			\cfence \ldots
	}

	\trans{\refsto}{\reflaw{swap-order}}
	\step{
		h \asgn Head \cbef \fence \cbef 
			return \asgn tasks[h \mod L] 
			\bef
			t \asgn Tail \cbef
	}
	\step{
		\qquad\qquad
		\guard{h < t} \cbef
			\cfence \ldots
	}

\end{derivation}
The access to local variable $h$ in the index expression now precedes the load $t \asgn Tail$.  Hence, if $h = t = 0$ initially, the deque is empty, and the
assignment $return \asgn tasks[h \mod L]$ sets $return$ to be the value at $tasks[0]$, which is an irrelevant value.  Now a sibling process may execute a
$put(v)$ and insert $v$ at $tasks[0]$, after it has been read, and increment $Tail$.  Then the original process resumes, reading $t \asgn 1$, and succeeding
the guard condition, eventually returning the default value at $tasks[0]$.
The trailing $\cfence$ in the original algorithm has no effect on this possible reordering as it occurs after the load.

%Note that the code is written at a higher level than direct ARM assembler; we assume that it is valid to translate it as we have done.

Swapping the order of this second $\cfence$ with the load of \code{task} eliminates the above reordering, and our analysis did not reveal any other problems.  In addition,
eliminating the first $\cfence$ does not change the possible outcomes.  The original placement of the control fence is unusual in that it comes after a load
and before an atomic load and store.  It is reasonable to assume that a \code{CAS} cannot be reordered before a branch since it involves a store to a global
address.  Therefore the placement of the control fence may indicate a minor misunderstanding of the subtleties of where a control fence must be placed to have the
desired effect; certainly, a control fence appears to be required for the algorithm to work correctly.

We tested 5 combinations of the three (modified) operations in parallel (as well as testing single operations and combinations of \code{put} and \code{take} on a single
thread).  The longest test to complete was a \code{push} with two \code{steal}s, which required 2 billion rewrites and 35 minutes.  
%A version with two \code{put}s on one thread in
%parallel with two \code{steal} operations did not complete within a reasonable timeframe.

\section{Related work}
\label{related-work}

%\paragraph{Alglave, Sewell, et al}
This work builds on a significant body of work in elucidating the behaviour of weak memory models in TSO, ARM and POWER via both operational and axiomatic semantics
\citep{x86-TSO,HerdingCats,UnderstandingPOWER,AxiomaticPower,ModellingARMv8}.  Those semantics were developed and validated through testing on real hardware and in consultation with
processor vendors.  We therefore had the easier task of validating our semantics against their results, in the form of the results of litmus tests.  The intention of
that body of work was to provide the foundation for higher-level verification of the sort that we have presented here.

More specifically, our model of the storage subsystem is similar to that of the operational models of \citep{UnderstandingPOWER,ModellingARMv8}.  However our thread
model is quite different, being defined in terms of relationships between actions.  The key difference is how we handle branching and the effects of speculative
execution.  The earlier models are complicated in the sense that they are closer to the real execution of instructions on a processor, involving restarting reads
if an earlier read invalidates the choice taken at a branch point.  We instead use a more abstract formulation of branches as guards.  Because speculative
execution should have no effect if an incorrect choice is made, it is straightforward to eliminate such behaviours.  However, the behaviour where the correct
choice is (eventually) made contains that choice as a guard action, before which later actions may have been reordered (if allowed by the rules of the
architecture).  Our semantics is presented in a conventional operational semantics style, where actions appear in the trace.  Unlike Plotkin-style operational
semantics \citep{Plotkin81,Plotkin}, we do not keep the state in the configurations, but as a first-order command of the language.  This style interacts well with
syntax-specific behaviours such as distinguishing between behaviours for registers or shared variables.  A similar approach is used by Owens \citep{OwensOcamllight}
and Abadi \& Harris \citep{AbadiHarris2009} in operational semantics; using the syntax of labels is also used in a denotational semantics by Brookes
\citep{BrookesSepLogic07}.

Our approach to modelling the non-multicopy atomic storage subsystem state is based on that of the operational model of \cite{UnderstandingPOWER}.  However, that model maintains several
partial orders on operations reflecting the nondeterminism in the system, whereas we let the nondeterminism be represented by choices in the operational rules.
This means we maintain a simpler data structure, a single global list of writes.

The axiomatic models, as exemplified by \citep{HerdingCats}, define relationships between instructions in a whole-system way, including relationships between
instructions in concurrent threads.  This gives a global view of how an architecture's reordering rules (and storage system) interact to reorder instructions in a
system.  Such global orderings are not immediately obvious from our pair-wise orderings on instructions.  On the other hand, those globals orderings become quite
complex and obscure some details, and it is unclear how to extract some of the generic principles such as \refeqn{reordering-principle}.  

We don't distinguish between ARM and POWER, and our model is less accurate to POWER with respect to the litmus tests than it is to ARM.  The difference between ARM
and POWER is accounted for in \HC
by weakening the POWER model to obtain their ARM model,
loosening the ``\T{po-loc}'' constraint, i.e., allowing loads and stores to the same variable on a process to not necessarily occur
in program order.  
This is fundamentally against our basic principles for reordering and we can't directly represent the same change in our framework.  
However, many of the behaviours discussed in \HC as being peculiar to ARM
were modelled by allowing forwarding (and eliminating earlier writes).

The model checking approach we developed exposed a bug in an algorithm in \citep{LeWorkStealingPPoPP13} in relation
to the placement of a control fence.  That paper includes a formal proof of the correctness of the algorithm based on the axiomatic model of
\citep{AxiomaticPower}.  The possible traces of the code were enumerated and validated against a set of conditions on adding and removing elements from the
deque (rather than with respect to an abstract specification of the deque).  As shown via derivations in \refsect{chaselev-verification} the reordering is
straightforward to observe directly by looking at the code. The reordering relation for the ARM architecture show that the first control fence is redundant
(does not prevent any reorderings) because it can be reordered to the previous fence. Similarly the load in the branch can come before the branch point
itself (speculatively), and hence before the earlier load controlling the branch. The control fence, in its original position, does nothing to prevent this.
The semantics of \cite{HerdingCats} does not uncover this anomaly as directly because it is more complex to construct the whole-code relations, 
while operational
models that are more closely based on hardware mechanisms \cite{UnderstandingPOWER,ModellingARMv8} are more complex and obscure this relatively straightforward property.

\paragraph{Other approaches}
Describing weak memory models has been tackled in a variety of other approaches.  Our results agree broadly with those of \citep{Lahav2016} in that many reorderings
cannot be explained locally only, and need a storage system for explanation.  That work provides some results relating TSO, ARM and POWER, although that model does
not handle control fences.  Alglave and Cousot \citep{AlglaveCousot2017} develop an Owicki-Gries style proof method for concurrent algorithms in which the program
text is annotated with invariants.  The algebraic approach we adopt in is closer to the style of the Concurrent Kleene Algebra \citep{CKA},
where sequential and parallel composition contribute to the event ordering.

\paragraph{Tool support and model checking}
The tool we developed was written in Maude without any specific attempt to specialise for the performance issues of weak memory models.  The number of interleavings
of parallel processes is factorial in the total number of actions, and this explosion is compounded by local reorderings.  The relatively simple algorithms in
\refsect{higher-level-code} became infeasible to model check for 4 or more parallel processes.  Potentially we could restructure the semantics to develop partial orders on
actions rather than traces, and hence benefit from other model checking approaches such as \citep{PartialOrderChecking}.
However, our tool does perform well on the litmus tests in comparison with the tools in \citep{UnderstandingPOWER,ModellingARMv8}, which did not report results for some
tests that our tool was able to.  
%The corresponding hardware that was used may be a factor in this, however.

\section{Conclusion}

We have built upon earlier work to devise a model of relaxed memory which is relatively straightforward to define and extend, and which lends itself to model checking
and formal analysis.  While abstracting away from the details of the architecture, we believe it provides a complementary insight into why some reorderings are
allowed, requiring a pair-wise relationship rather than system-wide.  
%It remains to be seen if this pair-wise approach fits with the C11 memory model approach
%\citep{BoehmAdveC++Concurrency,HerdingCats,VafeiadisC11} where individual instructions are labelled with an ordering type; our research into that memory model is
%ongoing.  

%Because the behaviour of commercial hardware is typically protected, it is not possible to conform to a single formal specification.  Fortunately the research of
%Alglave et al.{} has been as thorough as possible, including running thousands of litmus tests millions or billions of times on real hardware, collecting the results
%against formal conditions, working with a representative of ARM, and making the results of the hardware and model tests easily available.  Without
%access to thi resource we could not have built our own semantics.

We have described the ordering condition as syntactic constraints on atomic actions.  This fits with the low-level decisions of hardware processors such as ARM and
POWER, but variable references are not in general maintained by compilers (for instance, $r \asgn y \times 0$ may be reduced to $r \asgn 0$, eliminating what is
syntactically a load).  Our main reordering principle \refeqn{reordering-principle} is based on semantic concerns: preserving sequential behaviour.  As such our
semantics may be applicable as a basis for understanding
the interplay of software memory models, compiler optimisations and hardware memory models \citep{PromisingSemantics}.

One of the key aspects of our semantics is that it uses labels to describe the traces.  Traditional Plotkin-style operational semantics \citep{Plotkin}
keep the state in the configuration of the rule, which makes it difficult, if not impossible, to determine the exact nature of an instruction executed
by a subterm, and hence to check
the constraints of the reordering principle \refeqn{reordering-principle}.  In addition, the use of guards from Dijkstra's guarded command language
\citep{GuardedCommands} allowed an abstract treatment of speculative execution and the effect of early loads (loads occurring before a branch has been evaluated).
Systems are specified in a term structure \refeqn{system-structure} which theoretically lends itself to algebraic manipulation,
with the intention of being able to formally prove correctness of higher-level algorithms.  The first
steps towards this goal are taken in \refsects{treiber-verification}{chaselev-verification}.
%We may improve on the reachability model checking technique presented there to instead check for linearisability directly, by explicitly including invoke/response
%pairs for data structure operations and determining if they may be reordered to match a history from an abstract object.

%\noindent{\bf Acknowledgements} 
\paragraph{\bf Acknowledgements} 
We thank Kirsten Winter and Ian Hayes for feedback on this work.  
We also thank Luc Maranget,
Peter Sewell, Jade Alglave, and Christopher Pulte
for assistance with litmus test analysis.
The work was supported by Australian Research Council Discovery Grant DP160102457.
%\end{acks}

%\bibliographystyle{plain}
%\bibliographystyle{ACM-Reference-Format}
\bibliographystyle{elsarticle-num}
%\citestyle{acmauthoryear}

{
\raggedright
\bibliography{biblio,colvinpubs}
}

\appendix

\section{Shadow registers}
\label{shadow-registers}

Consider the test \T{MP+dmb+rs} from \cite{ModellingARMv8}\footnote{\url{
http://www.cl.cam.ac.uk/~pes20/arm-supplemental/src/MP+dmb+rs.litmus
}}
\begin{align}
	(x \asgn 1 \cbef \fence \cbef y \asgn 1)
	~~ \pl ~~
	(r_1 \asgn y \cbef r_2 \asgn r_1 \cbef r_1 \asgn x)
\end{align}
Shadow registers allow behaviour where the above process
can reach
a final state where $r_1 = 0$ and $r_2 = 1$.
This implies that $r_1$ has read the initial value of $x$ after reading an updated version of $y$, which is prohibited by the fence in the first
process.
According to \refeqn{reordering-principle} reordering of any of the register operations in process 2 should be disallowed because there is a dependency (each references $r_1$).  But
it seems that the first two are collapsed into a load $r_2 \asgn y$, and this can be reordered with $r_1 \asgn x$ as there is no dependency.  This
would be sound except that the value of $r_1$ should be preserved.  This litmus test and surrounding discussion indicate that the final value of
shadow registers should not be referenced, and that shadow registers are used for storing temporary calculated values.  In this case, we do not need to
model such registers for high-level code as they do not have the typical semantics of a local variable (one would instead use an explicit temporary
variable rather than reuse a variable name such as $r_1$).

While we rule out shadow registers from consideration, we could straightforwardly extend our framework to allow them.  This would require distinguishing them as a
special variable type with tailored instruction types and liberal reordering relation.  This change would be unlikely to affect
the results of any other litmus test, since reusing registers does not typically occur; nor would the change be likely to affect the majority of high-level code
since, as discussed above, local variable names are typically not reused.  However, these changes would violate the principle of sequential consistency for the
local thread \refeqn{reordering-principle}.  Hence we consider shadow registers to be a special type of variable with a special semantics designed for low-level
use, and therefore believe that their behaviour does not invalidate the principles of our reordering model.

\end{document}